\documentclass[sigconf]{acmart}
\AtBeginDocument{%
  \providecommand\BibTeX{{%
    \normalfont B\kern-0.5em{\scshape i\kern-0.25em b}\kern-0.8em\TeX}}}

\setcopyright{acmcopyright}
\copyrightyear{2018}
\acmYear{2018}
\acmDOI{XXXXXXX.XXXXXXX}
\acmPrice{15.00}
\acmISBN{978-1-4503-XXXX-X/18/06}

\usepackage[T1]{fontenc}
\usepackage{graphicx}
\usepackage{lineno}
\usepackage{balance}

\usepackage{flushend}
\usepackage{dcolumn}
\usepackage{bm, color}
\usepackage{graphicx}
\usepackage{diagbox} 
\usepackage{subfig}
\usepackage{caption}
\usepackage{algpseudocode}
\usepackage{algorithmicx}
\usepackage{algorithm}
\usepackage{makecell}
\usepackage{multirow}
\usepackage{hyperref}
\usepackage{smartdiagram}
\usepackage{verbatim}
\hypersetup{
    colorlinks=true,
    linkcolor=blue,
    filecolor=blue,      
    urlcolor=blue,
    citecolor=blue,
}

\newtheorem{definition}{Definition}
\newtheorem{corollary}{Corollary}
\newtheorem{theorem}{Theorem}
\newtheorem{lemma}{Lemma}
\newtheorem{proposition}{Proposition}
\newtheorem{problem}{Problem}
\newtheorem{remark}{Remark}

\usepackage{quantikz}

\usepackage{tikz}

\usetikzlibrary{positioning,patterns}
\usetikzlibrary{decorations.pathreplacing,decorations.pathmorphing}
\colorlet{colred}{red!22}
\colorlet{colcyan}{cyan!22}
\colorlet{colV}{blue!40}
\colorlet{colBorder}{gray!70}
\tikzset
  {mybox/.style=
    {rectangle,rounded corners,drop shadow,minimum height=1cm,
     minimum width=2cm,align=center,fill=#1,draw=colBorder,line width=1pt
    },
   myarrow/.style=
    {draw=#1,line width=3pt,-stealth,rounded corners
    },
   mylabel/.style={text=#1}
  }
\newtheorem{example}{Example}

\renewcommand{\epsilon}{\varepsilon}
\newcommand{\functiondot}{\,\mathord{\cdot}\,}
\newcommand{\bigO}[1]{\mathcal{O}(#1)}

\newcommand{\setnocond}[1]{\{#1\}}

\renewcommand{\dh}{\ensuremath{\mathcal{D(H)}}}
\newcommand{\doo}{\ensuremath{\mathcal{D(O)}}}

\renewcommand{\u}{\ensuremath{\mathcal{U}}}
\newcommand{\e}{\ensuremath{\mathcal{E}}}
\newcommand{\f}{\ensuremath{\mathcal{F}}}

\newcommand{\h}{\ensuremath{\mathcal{H}}}
\newcommand{\s}{\ensuremath{\mathcal{S}}}
\renewcommand{\i}{\ensuremath{\mathcal{I}}}

\renewcommand{\o}{\ensuremath{\mathcal{O}}}
\newcommand{\m}{\ensuremath{\mathcal{M}}}

\newcommand{\nn}{\ensuremath{\mathbb{N}}}

\newcommand{\ee}{\ensuremath{\mathbb{E}}}

 \renewcommand{\t}{\ensuremath{\mathcal{T}}}

 \newcommand{\tr}{{\rm tr}} 

\newcommand{\ketbra}[2]{| #1 \rangle \langle #2 |}

\newcommand{\n}{\ensuremath{\mathcal{N}}}

\newcommand{\dep}{\text{Dep}}
\newcommand{\gad}{\text{GAD}}
\copyrightyear{2025}
\acmYear{2025}
\setcopyright{acmlicensed}
\setcctype{by}
\acmConference[CCS '25]{Proceedings of the 2025 ACM SIGSAC Conference on Computer and Communications Security}{October 13--17, 2025}{Taipei, Taiwan}
\acmBooktitle{Proceedings of the 2025 ACM SIGSAC Conference on Computer and Communications Security (CCS '25), October 13--17, 2025, Taipei, Taiwan}\acmDOI{10.1145/3719027.3765178}
\acmISBN{979-8-4007-1525-9/2025/10}
\begin{document}

\title{Optimal Mechanisms for Quantum Local Differential Privacy}
\author{Ji Guan}
\orcid{0000-0002-3490-002}
\affiliation{%
\institution{Key Laboratory of System Software (Chinese Academy of Sciences) and State Key Laboratory of Computer Science, Institute of Software, Chinese Academy of Sciences,}
\city{Beijing}
\state{}
\country{China}
\postcode{100190}}
\email{guanji1992@gmail.com}
\renewcommand{\shortauthors}{J. Guan}
\begin{abstract}
Centralized differential privacy has been successfully applied to quantum computing and information processing to protect privacy and avoid leaks in the connections between neighboring quantum states. Consequently, quantum local differential privacy (QLDP) has been newly proposed to preserve quantum data privacy akin to the classical scenario where all states are viewed as neighboring states.  However, the exploration of the QLDP framework is still in its early stages, primarily conceptual, which poses challenges for its practical implementation in safeguarding quantum state privacy.

This paper delves into optimal QLDP mechanisms to balance privacy and utility to enhance the practical use of the QLDP framework. QLDP utilizes a parameter $\epsilon$ to manage privacy leaks and ensure the privacy of individual quantum states. The optimization of the QLDP value $\epsilon$, denoted as $\epsilon^*$, for any quantum mechanism is addressed as an optimization problem. The introduction of quantum noise is shown to provide privacy protections similar to classical scenarios, with quantum depolarizing noise identified as the optimal unital privatization mechanism within the QLDP framework. Unital mechanisms represent a diverse set of quantum mechanisms that encompass frequently employed quantum noise types. Quantum depolarizing noise optimizes both fidelity and trace distance utilities, which are crucial metrics in the field of quantum computation and information, and can be viewed as a quantum counterpart to classical randomized response methods. The study further explores the trade-off between utility and privacy across different quantum noise mechanisms, including unital and non-unital quantum noise mechanisms, through both analytical and numerically experimental approaches. This highlights the optimization of quantum depolarizing noise in the QLDP framework.
\end{abstract}

\ccsdesc[500]{Security and privacy~Formal methods and theory of security}
\ccsdesc[500]{Theory of computation~Quantum computation theory}

\keywords{Quantum Computing, Local Differential Privacy, Optimal Mechanism, Quantum Noise.}

%
\maketitle

\section{Introduction}
Quantum computers have the potential to solve specific problems significantly faster than classical computers through the utilization of quantum parallelism in the development of quantum algorithms~\cite{nielsen2010quantum}. Examples include Grover's algorithm~\cite{grover1996fast} for searching unstructured databases, Shor's algorithm~\cite{shor1994algorithms} for identifying prime factors of integers, and the HHL algorithm~\cite{harrow2009quantum} by Aram Harrow, Avinatan Hassidim, and Seth Lloyd for solving linear equations systems. This advancement could bring about revolutionary changes in various fields such as financial analysis~\cite{orus2019quantum}, drug discovery~\cite{wang2023recent}, and machine learning~\cite{biamonte2017quantum}. On another note, principles of quantum mechanics like the no-cloning theorem and the observer effect establish inherently secure communication channels~\cite{nielsen2010quantum}. Any unauthorized attempt to intercept quantum signals would inevitably disrupt them, thereby alerting the legitimate parties. This feature allows for the secure distribution of encryption keys, exemplified by the BB84 quantum key distribution (QKD) protocol~\cite{bennett2014quantum}. As quantum hardware techniques rapidly advance, quantum computers have been developed to exhibit quantum supremacy (outperforming classical computation), as seen in Google's \textit{Sycamore}~\cite{arute2019quantum}, USTC's \textit{Jiuzhang}~\cite{zhong2020quantum}, and \textit{Zhucongzhi}~\cite{zhu2022quantum}. Additionally, various quantum communication channels and networks have been demonstrated or deployed mainly for research and experimental purposes. Notable examples include the \textit{Chicago Quantum Exchange} in the United States, China's \textit{Micius Satellite}, Japan's \textit{Tokyo QKD Network}, and the Netherlands' \textit{Quantum City Internet}.

However, just like their classical counterparts, quantum computers and channels are susceptible to data privacy leakage. This is due to the various threats they encounter, including internal and external attacks like coherent attacks~\cite{furrer2012continuous}, entangling attacks~\cite{azuma2018entangling}, higher-energy state attacks~\cite{xu2023securing}, and quantum-side-channel attacks~\cite{xu2023exploration}. Some of these attacks are unique to quantum systems and necessitate new methods to safeguard against them, surpassing the capabilities of traditional classical defense mechanisms. Consequently, safeguarding the privacy of information stored in quantum states for computing and communication purposes presents significant challenges.

In the realm of classical computing, differential privacy stands out as a key method for safeguarding individual data~\cite{dwork2014algorithmic}. It is commonly divided into two categories: centralized differential privacy, where institutions publish databases containing information about multiple individuals or respond to queries based on such databases (e.g., the US Government releasing census data or companies like Netflix sharing proprietary data for testing advanced machine learning techniques), and local differential privacy, where individuals reveal personal details, such as voluntarily on social networking platforms. The former aims to safeguard data privacy against the interrelation between datasets, while the latter seeks to protect all information pertaining to an individual. Previous studies have endeavored to expand centralized differential privacy into the realm of quantum computing by establishing various meaningful relationships between neighboring quantum data sources~\cite{aaronson2019gentle,hirche2023quantum,angrisani2023unifying}. A more adaptable framework has recently been put forth~\cite{nuradha2023quantum} by building upon the classical concept of pufferfish privacy~\cite{kifer2014pufferfish}. Nevertheless, the practical implementation of quantum databases remains challenging in the current era of Noisy Intermediate-Scale Quantum (NISQ) computing. Presently, quantum systems are predominantly utilized in a distributed fashion. In the realm of quantum computing, users typically access quantum computers via cloud platforms (e.g., IBM's Qiskit, Amazon's Braket, and Google's Cirq), enabling them to remotely submit their data for computational tasks on these quantum machines. On the other hand, in quantum communication, the transfer and sharing of quantum states occur between remote user nodes within a communication network. Consequently, the deployment of quantum computing has predominantly focused on the local model where quantum systems gather quantum states directly from users as computational and communication resources, thereby emphasizing the critical need to protect users' information comprehensively. Therefore, integrating privacy measures into the quantum state at the client end before collection is essential to preserve individual data. Local differential privacy emerges as one of the most effective privacy measures in the context of local models. Despite its efficacy in classical environments, the potential of local differential privacy in quantum systems remains underexplored and is still in the conceptual stage. The notion of \emph{quantum local differential privacy (QLDP)} is introduced in~\cite{hirche2023quantum} as an extension of quantum centralized differential privacy, viewing all quantum states as neighboring states akin to the classical scenario. Certain basic properties of QLDP can be derived from prior studies on quantum centralized differential privacy \cite{hirche2023quantum} and the more adaptable framework --- quantum pufferfish privacy \cite{nuradha2023quantum}. However, this approach does not yet provide a practical QLDP framework for ensuring quantum state privacy in the current local model.

In this paper, we enhance the practical use of QLDP frameworks by exploring the optimal QLDP mechanisms, which balance privacy and utility. The key inquiries addressed are:

\begin{enumerate}
    \item How can we evaluate and manage the (optimal) local differential privacy of quantum mechanisms?
    \item How can we maintain a balance between preserving utility and safeguarding privacy in quantum systems?
\end{enumerate}

To tackle these issues, we commence by defining local differential privacy for quantum computers and communication systems based on
the classical counterpart’s definition, which aligns with the definition presented in~\cite{hirche2023quantum} from
a quantum information standpoint. Similar to the classical scenario, QLDP incorporates a parameter denoted as $\epsilon$ that governs the formal evaluation of privacy loss. We present a method for quantifying the optimization (minimum) of $\epsilon$, denoted as $\epsilon^*$, for any given quantum mechanism by solving an optimization problem. In order to ensure privacy, a post-processing theorem is demonstrated to ensure that clients' quantum state privacy is preserved through quantum noise mechanisms before aggregation, akin to the classical setting. This post-processing theorem is directly deduced from previous research on the post-processes of quantum centralized differential privacy~\cite{hirche2023quantum} and pufferfish privacy~\cite{nuradha2023quantum} by considering all quantum states as neighboring states.  

Subsequently, we analyze the $\epsilon^*$ value of different standard quantum noise mechanisms for practical application. It is noted that certain commonly employed quantum noise mechanisms do not yield a measurable QLDP value $\epsilon^*$ ($\epsilon^*<\infty$). The key issue at hand is the selection of a suitable mechanism that can provide a finite QLDP value $\epsilon^*$ as a minimum requirement. To tackle this challenge, we propose a method to identify efficient quantum noise mechanisms that result in a measurable QLDP value.

Moreover, it is noteworthy that all quantum mechanisms involve trade-offs between utility and privacy, and a primary objective of practical use in differential privacy is to optimize this balance. By considering two key quantum utility measures --- fidelity and trace distance --- we establish that the (quantum) depolarizing noise mechanism is the optimal choice, enhancing both utility measures, for all $\epsilon$-QLDP unital quantum mechanisms. The collection of unital quantum mechanisms encompasses a diverse range of quantum processes, such as the frequently employed quantum noises --- bit flip and phase flip, which will be elaborated on later. The depolarizing noise mechanism can be seen as a quantum adaptation of the classical differential privatization mechanism known as randomized response~\cite{warner1965randomized}, which optimizes the KL divergence utility~\cite{kairouz2016extremal}.


Last but not least, to validate the optimization of depolarizing noise mechanisms in terms of $\epsilon$-QLDP, we conduct both analytical and numerical analysis illustrating the trade-offs between utility and QLDP privacy across a range of standard quantum noise mechanisms, including unital noises (bit flip, phase flip, bit-phase flip, and depolarizing noises) as well as non-unital noises like (generalized) amplitude damping and phase damping~\cite{nielsen2010quantum}.

In summary, our main contributions are:
\begin{enumerate}
    \item  \textit{Offering} a practical approach to quantifying the optimal parameter $\epsilon^*$ of QLDP for any quantum mechanism by addressing an optimization problem.
    \item \textit{Introducing} an effective method to pinpoint efficient quantum noise mechanisms that yield a measurable QLDP value. 
    \item  \textit{Proving} that the depolarizing quantum noise mechanism is the most effective privatization mechanism, optimizing both fidelity and trace distance utilities overall  $\epsilon$-QLDP unital quantum mechanisms.
    \item \emph{Conducting} both analytical and numerical analyses to evaluate the balance between utility and privacy across different standard quantum noise mechanisms, emphasizing the superiority of the depolarizing noise mechanism.
\end{enumerate}

\subsection{Related Works}
{\bf Quantum Centralized versus Local Differential Privacy.} In recent years, there has been successful progress in extending centralized differential privacy to quantum computing and information processing to enhance privacy and prevent leaks in neighboring relationships of quantum states. These neighboring relationships are defined qualitatively and quantitatively using informatic measures --- local operation~\cite{aaronson2019gentle} and trace distance~\cite{guan2023detecting}, respectively. The former involves monic classical neighboring through a local operation to transfer a state to another, while the latter measures the distance between quantum states on a scale from 0 to 1. Moreover, the evaluation and formal verification of quantum centralized differential privacy have been established~\cite{guan2023detecting}, along with the establishment of a connection to quantum gentle measurements~\cite{aaronson2019gentle}.

On the other hand, quantum local differential privacy (QLDP), a generalization of LDP, was introduced to ensure that two distinct quantum states passed through a quantum mechanism are indistinguishable by any measurement~\cite{hirche2023quantum} to maintain quantum state privacy. While previous studies have delved into the applications of the QLDP framework in quantum statistical query models~\cite{angrisani2022quantum} and quantum hypothesis testing~\cite{nuradha2024contraction,angrisani2025quantum}, there has been limited direct research on effectively implementing the QLDP approach to safeguard quantum state privacy while preserving utility. This paper aims to measure and describe the balance between the utility and privacy of unital quantum privacy mechanisms. Our findings indicate that the depolarizing noise mechanism, a quantum generalization of the classical randomized response mechanism, emerges as the optimal solution for managing this balance effectively.

{\bf Classical versus Quantum Local Differential Privacy.} The concept of classical local differential privacy (LDP) has been extensively researched and implemented in real-world applications like Google's Chrome and Apple's iOS keyboard to safeguard user privacy. However, the methods used in classical LDP cannot be directly translated to the quantum realm, where QLDP emerges as an extension of LDP. Verifying that a quantum mechanism adheres to QLDP is equivalent to confirming that an infinite number of classical randomized mechanisms comply with LDP (refer to Lemma~\ref{lem:QLDP_LDP}). Therefore, novel techniques, such as eigenvalue analysis of quantum mechanisms, are being developed in this paper to explore the fundamental properties of QLDP and ensure its practical efficacy in preserving the privacy of quantum states. 

{\bf Quantum Divergence-based versus Eigenvalue-based Techniques.} In the realm of quantum centralized and local differential privacy, previous research has predominantly relied on techniques stemming from quantum information theory, particularly focusing on quantum divergences like the quantum hockey-stick divergence and Datta–Leditzky divergence~\cite{hirche2023quantum}. Quantum differential privacy is akin to classical privacy in that it can be viewed through the lens of quantum divergences. One notable advantage of this approach is that quantum differential privacy is determined solely based on the output states of a quantum mechanism, eliminating the need for verification after each quantum measurement. By leveraging the known characteristics of quantum divergences, such as contractility and subadditivity, post-process theorems have been established and composition theorems have been built to address the parallel composability for product and classically correlated quantum states within the framework of quantum centralized differential privacy~\cite{hirche2023quantum,nuradha2023quantum}. However, while quantum divergence operates with specific quantum states, quantum differential privacy extends to any pair of (neighboring) quantum states within a system, making it challenging to formulate a closed characterization for quantum differentially private mechanisms. To address this limitation, a novel technique based on the maximum and minimum eigenvalues of quantum mechanisms across pure states has been introduced in this paper to analyze quantum differential privacy. This approach enables the optimization of the QLDP value, denoted as $\epsilon^*$, as a solution to an optimization problem, 
establishing a balance between utility and privacy in terms of differential privacy for quantum mechanisms. Consequently, identifying depolarizing quantum noise as the optimal mechanism among all unital quantum mechanics offers a practical framework for enhancing quantum state privacy within the QLDP context.
\section{Preliminaries}
In this section, we aim to present the fundamental concepts of quantum computing in a more accessible manner for the reader by using mathematical explanations.

Quantum computing, in essence, represents a novel circuit model that departs from traditional classical circuits. It offers computational advantages and employs quantum communication protocols that utilize quantum channels as opposed to classical ones for transmitting quantum states. Both quantum circuits and channels operate based on {quantum states} as inputs, enabling them to carry classical information essential for executing classical computational tasks and transmitting classical messages. The information encoded within these quantum states can be decoded through {quantum measurement} processes. Therefore, the execution of comprehensive quantum computational or communication tasks involves three key components: \emph{quantum states, quantum mechanisms (quantum circuits or channels)}, and \emph{quantum measurements}. In the subsequent sections, we will delve into each of these components individually.

\subsection{Quantum State} 
{\it Quantum Bit.} In the classical realm, a bit represents the fundamental unit of data characterized by two distinct states, $0$ and $1$. In quantum computing, a \emph{quantum bit (qubit)} is a 2-dimensional unit vector $\vec{q}$ existing within a linear complex vector space spanned by the computational basis $\{\vec{0}, \vec{1}\}$, defined as:
\[\vec{q}\coloneqq\left(\begin{matrix}
    a\\
    b
\end{matrix}\right )=a\left(\begin{matrix}
    1\\
    0
\end{matrix}\right )+b\left(\begin{matrix}
    0\\
    1
\end{matrix}\right )=a \vec{0}+b\vec{1} .\]
Here, $\vec{0}=\left(\begin{matrix}
    1\\
    0
\end{matrix}\right)$ and $\vec{1}=\left(\begin{matrix}
    0\\
    1
\end{matrix}\right)$, and $a$ and $b$ are complex numbers satisfying the normalization condition $|\vec{q}|\coloneqq\sqrt{aa^*+bb^*}=1$. A qubit can be interpreted as a probabilistic distribution over $\{\vec{0},\vec{1}\}$, corresponding to the classical states $\{0,1\}$. However, this distribution is not unique and is contingent upon the method of observation (specifically, the choice of basis). Further elucidation on this matter will be provided subsequently.

{\it Bra–ket Representation.}  In quantum computation and information, the bra-ket representation, also known as Dirac notation, is commonly employed to depict quantum states. This notation, utilizing angle brackets $\langle$ and $\rangle$ along with a vertical bar $|$, is utilized to describe linear algebra and linear operators on complex vector spaces and their dual spaces. It is specifically tailored to simplify calculations frequently encountered in quantum mechanics, hence its widespread application in this field.

In this notation, "bras" and "kets" are constructed using angle brackets and a vertical bar to represent a column vector and a row vector, respectively. For instance, in the case of a qubit, the representation is as follows:
\[\ket{x}\coloneqq\vec{x}=\left(\begin{matrix}
    a\\
    b
\end{matrix}\right )\quad \bra{x}\coloneqq\vec{x}^{\dagger}=\left(\begin{matrix}
    a^*,b^*
\end{matrix}\right ).\]
Here, $\vec{x}^{\dagger}$ denotes the complex conjugate and transpose of $\vec{x}$, and $a^*$ represents the conjugate of the complex number $a$.

For a concrete example, the computational basis of a qubit consists of $\ket{0}$ and $\ket{1}$ corresponding to the classical bit:
\[\ket{0}=\left(\begin{matrix}
    1\\
    0
\end{matrix}\right )\quad \ket{1}=\left(\begin{matrix}
    0\\
    1
\end{matrix}\right ). \]

These states, $\ket{0}$ and $\ket{1}$, can be combined to create larger quantum systems through their tensor product: $\ket{k,l}=\ket{k}\otimes \ket{l}$ for $k,l\in \{0,1\}$. In the case of a 2-qubit system, we have
\[\ket{0,0}=\left(\begin{matrix}
    1\\
    0\\
    0\\
    0
\end{matrix}\right ), \ket{0,1}=\left(\begin{matrix}
    0\\
    1\\
    0\\
    0
\end{matrix}\right),   \ket{1,0}=\left(\begin{matrix}
    0\\
    0\\
    1\\
    0
\end{matrix}\right), 
 \ket{1,1}=\left(\begin{matrix}
    0\\
    0\\
    0\\
    1
\end{matrix}\right ).\]
The tensor product of the computational basis $\ket{0}$ and $\ket{1}$ corresponds to the bit string of the classical bits $0$ and $1$.

{\it Pure Quantum State.} In accordance with the bra-ket notation, a state $\ket{\psi}$ comprising $n$ qubits can be constructed by combining individual qubits as shown below:
\begin{equation*}\label{Eq:state}
\ket{\psi}\coloneqq\left(
    a_0,
    a_1,
    \ldots,
    a_{2^n-1}
\right )^T \text{ and ensuring normalization } |\psi|=1,
\end{equation*}
where $|\psi|\coloneqq\sqrt{\braket{\psi}{\psi}}=\sqrt{\sum_{0\leq k\leq 2^n-1}a_{k}a_{k}^*}.$ Generally, $\ket{\psi}$ is referred to as a \emph{pure state}, distinct from \emph{mixed states} discussed later. Essentially, a pure state $\ket{\psi}$ consisting of $n$ qubits can be represented as a normalized column vector in a Hilbert space (a finite-dimensional linear space) $\h$ with a dimension of $2^n$. 

Moreover, $\ket{\psi}$ can be expressed as a sum over binary strings of length $n$ as follows:
\begin{equation*}
a_{0}\ket{0,0,\ldots,0}+a_{1}\ket{1,0,\ldots,0}+\cdots+ a_{2^n-1}\ket{1,1,\ldots,1}.
\end{equation*}
Here, each $a_k$ represents the \emph{amplitudes} of $\ket{\psi}$.

To facilitate the understanding of the reader, we summarize the linear algebra concepts pertaining to the pure state $\ket{\psi}$ utilized in this paper using the bra-ket notation as described below:
\begin{enumerate}
    \item $\ket{\psi}$ denotes an $n$-qubit unit complex column vector (quantum pure state) labeled as $\psi$;
    \item $\bra{\psi}\coloneqq \ket{\psi}^\dagger$ represents the complex conjugate and transpose of $\ket{\psi}$;
    \item $\braket{\psi_1}{\psi_2} \coloneqq \bra{\psi_1} \ket{\psi_2}$ signifies the inner product of $\ket{\psi_1}$ and $\ket{\psi_2}$;
    \item $\ketbra{\psi_1}{\psi_2} \coloneqq \ket{\psi_1}\cdot\bra{\psi_2}$ denotes the outer product of $\ket{\psi_1}$ and $\ket{\psi_2}$; specifically, $\psi\coloneqq\ketbra{\psi}{\psi}$ represents the outer product of $\ket{\psi}$ with itself.
\end{enumerate}

\textit{Quantum State Encoding.} In order to apply quantum circuits to solve practical classical problems and transfer classical information through channel mechanisms, the initial step involves encoding classical data into quantum states. Specifically, a classical unit vector $\vec{x}=(a_0,a_1,\cdots,a_{2^n-1})$ can be represented by the amplitudes of a quantum state $\ket{\psi}$, known as \emph{amplitude encoding}. For more general cases, including non-unit vectors, various encoding methods such as \emph{angle encoding} and \emph{bit encoding} can be employed~\cite{guan2023detecting}. These encoding techniques can be physically realized through quantum circuits, a process referred to as quantum state preparation.

{\it Mixed Quantum State.} In quantum mechanics, uncertainty is a common feature of quantum systems due to quantum noise and measurements. The concept of a \emph{quantum mixed state} $\rho$ is introduced to describe the uncertainty of possible pure quantum states. It can be expressed as
\begin{equation}\label{Eq:ensemble}
    \begin{aligned}
        \rho = \sum_{k} p_{k}\ket{\psi_{k}}\bra{\psi_{k}}.
    \end{aligned}
\end{equation}
Here, $\setnocond{(p_{k}, \ket{\psi_{k}})}_{k}$ denotes an ensemble, signifying that the quantum state is $\ket{\psi_{k}}$ with probability $p_{k}$. Moreover, if $\ket{\psi_k}$ are mutually orthogonal (i.e., $\braket{\psi_i}{\psi_j}=0$ for all $i\not = j$), then the decomposition of $\rho$ above represents the eigendecomposition with $\ket{\psi_{k}}$ as the eigenvector of $\rho$ corresponding to eigenvalue $p_{k}$. In the context of mixed quantum states, a pure quantum state $\psi$ can be seen as a mixed state $\psi=\ketbra{\psi}{\psi}$. To ensure clarity, in the subsequent discussion, the term "quantum states" will specifically denote mixed quantum states, given the broader context.

From a mathematical viewpoint, a mixed quantum state $\rho$  is  a $2^n$-by-$2^n$ matrix on a $n$-qubit system $\h$ that satisfies three conditions:
\begin{enumerate}
    \item \emph{Hermitian} $\rho^\dagger = \rho$. Here $\dag$ denotes the complex conjugate and transpose of matrices;
    \item \emph{positive semi-definite} $\bra{\psi} \rho \ket{\psi} \geq 0$ for all pure quantum state $\ket{\psi} \in \h$;
    \item \emph{unit trace} $\tr(\rho) = \sum_{k} \bra{\psi_k} \rho \ket{\psi_k} = 1$ defines the trace of $\rho$ as the total of the diagonal elements of $\rho$. Here, $\tr(\rho)$ signifies the trace of $\rho$, which is the sum of the diagonal elements of $\rho$, and $\{\ket{\psi_k}\}$ represents a unit basis of $\h$ that extracts the diagonal element of $\rho$.
\end{enumerate}

\subsection{Quantum Mechanism} 
Quantum mechanisms governed by quantum mechanics can be broadly categorized into two types: quantum circuits for computation and quantum channels for information transmission. Both quantum circuits and channels on a Hilbert space $\h$ are mathematically described by a \emph{super-operator} denoted as $\e$, which transforms an input quantum state $\rho$ into an output state $\rho'$ according to the equation:
\begin{equation}\label{Eq:evolution}
	\rho' = \e(\rho).
\end{equation}
The super-operator $\e(\functiondot)$ mathematically represents a linear mapping from $\dh$ to $\dh$, where $\dh$ is the set of all quantum states on $\h$. According to the \emph{Kraus representation theorem}~\cite{choi1975completely}, $\e$ can be represented by a finite set of matrices $\{E_{k}\}$ on $\h$ with the expression 
\[\e(\rho) = \sum_{k} E_k \rho E_{k}^{\dag} \qquad \forall \rho\in \dh.\] 
This representation also satisfies the trace-preserving condition, which is expressed as $\sum_{k} E_{k}^{\dag} E_{k} = I$, where $I$ is the identity matrix on $\h$. Furthermore, if $\e(I)=\sum_{k} E_{k} E_{k}^\dagger = I$, then $\e$ is a \emph{unital} quantum mechanism.

In classical differential privacy, the introduction of noise mechanisms is a primary method to safeguard privacy. Similarly, in the realm of quantum privacy, we can employ analogous techniques, which will be elaborated upon later. Below, we outline some typical and standard quantum noise mechanisms represented by super-operators. In contrast to the standard quantum mechanism denoted by $\e$, we specifically highlight the quantum noise mechanism represented by the symbol $\n$.
\begin{example}[Quantum Noise Mechanisms]\label{Exa:noise}  
Quantum computation and information fields commonly involve quantum noises on a single qubit~\cite{nielsen2010quantum}.  General quantum noise with multiple qubits can be constructed based on these individual noises. One type of 1-qubit noise involves Pauli matrices ($X, Y,$ and $Z$), which give rise to four types of errors in quantum systems similar to classical errors. These matrices consist of three 2-by-2 complex Hermitian matrices and together with the identity matrix $I$, they form a basis for 2-by-2 matrices.
    \[    
          X=\begin{bmatrix}
          0&1\\
          1&0
          \end{bmatrix}, \quad Y=\begin{bmatrix}
          0&-i\\
          i&0
          \end{bmatrix}, \quad Z=\begin{bmatrix}
          1&0\\
          0&-1
          \end{bmatrix}.
\]
By manipulating these matrices with a probability $0\leq p\leq 1$, four Pauli-type noise mechanisms can be created to control the introduction of errors, where errors occur with a probability of $1-p$ (quantum states are staying unchanged with a \emph{noiseless probability} $p$).
\begin{enumerate}
    \item X flip (Bit flip) noise mechanism:
    \[\n_{\text{XF}}(\rho)={p}\rho+(1-p)X\rho X\]
    using Kraus matrices $\{\sqrt{p}I, \sqrt{1-p}X\}$;
    \item Z flip (Phase flip) noise mechanism:
    \[\n_{\text{ZF}}(\rho)={p}\rho+(1-p)Z\rho Z\]
using Kraus matrices $\{\sqrt{p}I, \sqrt{1-p}Z\}$;
    \item Y flip (Bit–phase flip) noise mechanism:
    \[\n_{\text{YF}}(\rho)={p}\rho+(1-p)Y\rho Y\]
    using Kraus matrices $\{\sqrt{p}I, \sqrt{1-p}Y\}$;
    \item Depolarizing noise mechanism:
    \[\n_{\dep}(\rho)=\frac{1+3p}{4}\rho+\frac{1-p}{4}(X\rho X+Y\rho Y+Z\rho Z)\]
    using Kraus matrices $\{\frac{\sqrt{1+3p}}{2}I, \frac{\sqrt{1-p}}{2}X,\frac{\sqrt{1-p}}{2}Y,\frac{\sqrt{1-p}}{2}Z\}$.
    
    Alternative form of $\n_{\dep}$ is 
    \[\n_{\dep}(\rho)=p\rho+(1-p)\frac{I}{2}.\]
\end{enumerate}

Furthermore, a parameter denoted as $\gamma$ can be introduced to regulate the energy damping within a quantum system, with $\gamma$ representing the probability associated with energy loss. Subsequently, three distinct types of damping noise mechanisms are derived:
\begin{enumerate}
\item Phase damping noise mechanism:
\[\mathcal{N}_{\text{PD}}(\rho) = N_0 \rho N_0^\dagger + N_1 \rho N_1^\dagger\]
where the corresponding Kraus matrices are given by:
\[N_0 = \begin{bmatrix}
        1 & 0 \\
        0 & \sqrt{1-\gamma}
    \end{bmatrix}, \quad N_1 = \begin{bmatrix}
        0 & 0 \\
        0 & \sqrt{\gamma}
    \end{bmatrix}.\]
\item Amplitude damping noise mechanism:
\[\mathcal{N}_{\text{AD}}(\rho) = N_0 \rho N_0^\dagger + N_1 \rho N_1^\dagger\]
where the corresponding Kraus matrices are given by:
\[N_0 = \begin{bmatrix}
        1 & 0 \\
        0 & \sqrt{1-\gamma}
    \end{bmatrix}, \quad N_1 = \begin{bmatrix}
        0 & \sqrt{\gamma} \\
        0 & 0
    \end{bmatrix}.\]
\item Generalized amplitude damping noise mechanism:
\[\mathcal{N}_{GAD}(\rho) = N_0 \rho N_0^\dagger + N_1 \rho N_1^\dagger + N_2 \rho N_2^\dagger + N_3 \rho N_3^\dagger\]
where the Kraus matrices are defined as follows:
\[N_0 = \sqrt{q} \begin{bmatrix}
        1 & 0 \\
        0 & \sqrt{1-\gamma}
    \end{bmatrix}, \quad N_2 = \sqrt{1-q} \begin{bmatrix}
        0 & 0 \\
        \sqrt{\gamma} & 0
    \end{bmatrix}\]
\[N_1 = \sqrt{q} \begin{bmatrix}
        0 & \sqrt{\gamma} \\
        0 & 0
    \end{bmatrix}, \quad N_3 = \sqrt{1-q} \begin{bmatrix}
        \sqrt{1-\gamma} & 0 \\
        0 & 1
    \end{bmatrix}.\]
    Here, a probability $0 \leq q \leq 1$ represents the likelihood of Kraus matrices $\{N_0, N_1\}$ contributing to $\n_{\text{AD}}$. Specifically, when $q=1$, the generalized amplitude damping noise $\n_{\gad}$ is simplified to the standard noise model $\n_{\text{AD}}$.
\end{enumerate}
Through basic calculations, it is evident that amplitude damping and generalized amplitude damping noises are non-unital, whereas all other noises are unital.
\end{example}
\subsection{Quantum Measurement}\label{subsec:measurement}
To extract classical information from a quantum state $\rho$, the only method is to conduct a quantum measurement on $\rho$. In mathematical terms, a quantum measurement $\m$ is a stochastic mechanism defined over a finite set $\o$ of measurement outcomes: 
\begin{equation*}
\begin{aligned}
\m: \dh\rightarrow \doo\text{ with } \text{Pr}[\m(\rho)=k]=\tr(M_{k}\rho) \text{ }\forall \rho\in\dh.
 \end{aligned}
\end{equation*}
In this context, $\doo$ represents the set of probability distributions over the measurement outcomes set $\o$, $\{M_{k}\}_{k\in\o}$ denotes a collection of positive semi-definite matrices on the state (Hilbert) space $\h$, and $\text{Pr}[\m(\rho)=k]$ signifies the probability of observing the outcome $k$ through the measurement $\m$.
It is essential to note that the set of measurements $\{M_{k}\}_{k \in \o}$ adheres to the \emph{unity condition} $\sum_{k} M_{k}=I$, ensuring that the sum of probabilities of all outcomes equals 1. In simpler terms, $\sum_{k} \tr(M_{k} \rho) = \tr(\sum_{k} M_{k} \rho) = \tr(\rho) = 1$.

These types of measurements are commonly referred to as \emph{Positive Operator-Valued Measures} and are extensively utilized for determining outcome probabilities without involving the post-measurement quantum states. It is important to note that following the measurement, the state undergoes collapse or alteration based on the measurement outcome $k$, demonstrating a fundamental distinction from classical computation.




\begin{figure*}[!htp]
    \centering
    \includegraphics[width=\linewidth]{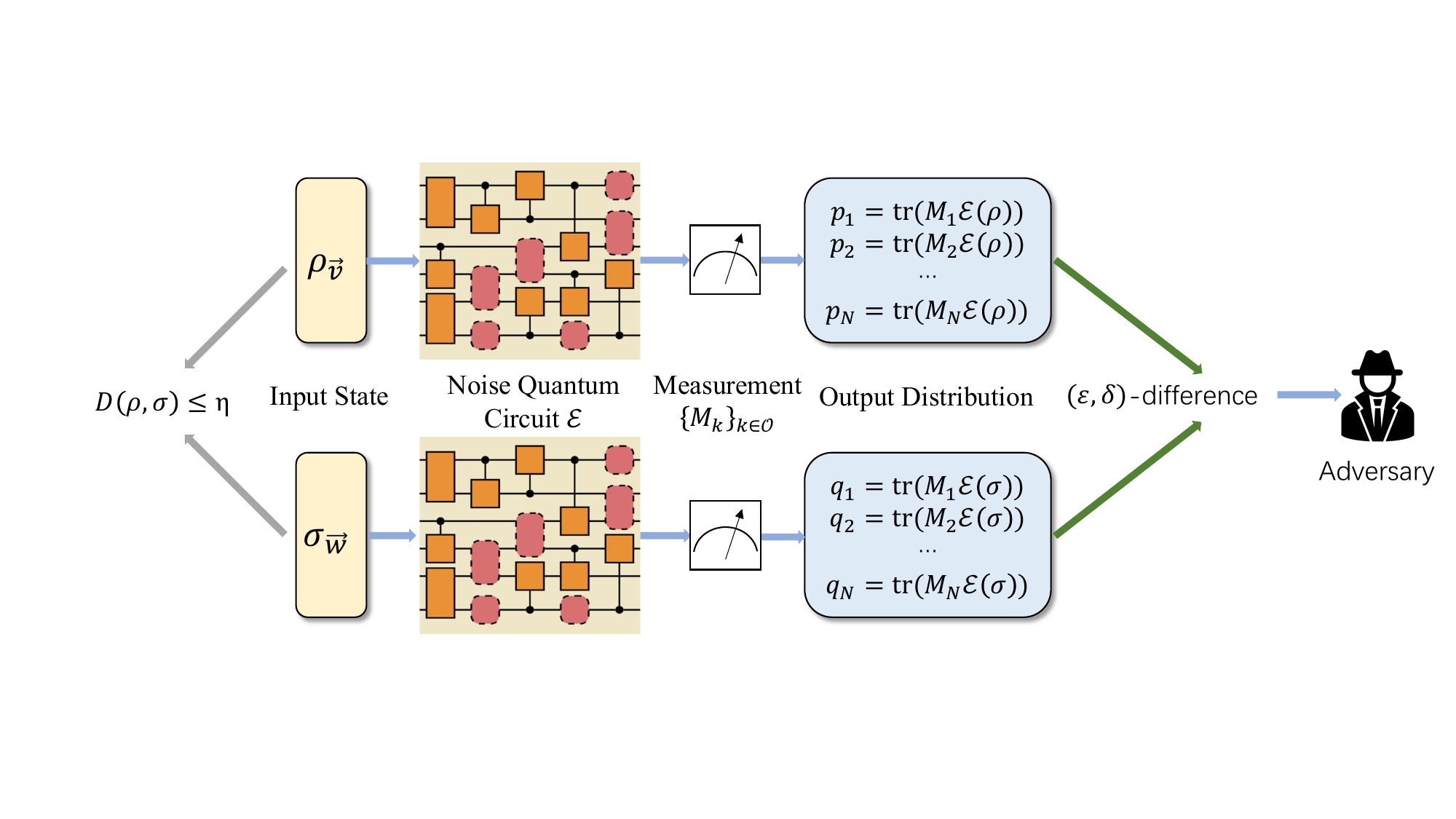}
    \caption{The framework of quantum local differential privacy}
    \label{fig:framework}
\end{figure*}

\section{Quantum Threat Model}\label{Sec:threat_model}
Creating a threat model for local differential privacy in quantum systems requires defining possible adversaries and outlining their abilities. This section aims to outline the QLDP framework's threat model, highlighting adversaries' distinctive characteristics in quantum environments. These adversaries can access quantum states directly and execute various quantum measurements. 

In order to enhance the resilience of our defense strategies, we are addressing \emph{ the (quantum) adversary with the ability to directly manipulate the quantum states published by quantum computers and communication channels and conduct diverse quantum measurements on them}, as illustrated in the right side of Fig.~\ref{fig:framework}. This is a departure from conventional attacks, where adversaries are limited to observing the outcomes of a specific quantum measurement only~\cite{guan2023detecting}. As discussed in the previous section, classical defense mechanisms can effectively thwart such attacks due to the stochastic nature of the measurement process. However, in our more stringent scenario, the adversary can execute any quantum measurement on the state acquired from the quantum system and infer the state's confidentiality. Consequently, there is a necessity to introduce quantum defense measures to uphold local differential privacy in quantum systems, given the infinite potential quantum measurements for a quantum system. In this paper, we use quantum noise mechanisms to protect QLDP and advocate for depolarizing noise as the optimal choice to strike a balance between utility and privacy.

\section{Quantum Local Differential Privacy}
In this section, we begin by explaining local differential privacy for quantum systems based on the classical counterpart's definition and examining its connection with classical local differential privacy through quantum measurements. Following that, we demonstrate the assessment of the local differential privacy in quantum mechanisms and introduce a post-processing theorem for leveraging quantum noise to safeguard the privacy of the quantum state. For the convenience of the reader, we put all proofs of theoretical results in the appendix of the full version of our paper~\cite{guan2024optimal}. 

In accordance with the threat model we face in Section~\ref{Sec:threat_model}, the attacker has the ability to access the quantum state and conduct any quantum measurement on it. To safeguard the quantum state privacy under such conditions, quantum local differential privacy is introduced as follows. 

\begin{definition}\label{Def:QLDP}
A quantum mechanism $\e$ operating on a Hilbert space $\h$ is $\epsilon$-\emph{quantum locally differentially private ($\epsilon$-QLDP)} if, for any quantum measurement $\m=\{M_{k}: k\in\o_{\m}\}$ and any quantum states $\rho,\sigma\in\dh$, the following inequality holds:
\begin{equation*} 
\text{Pr}[\m(\e(\rho))\in\s]\leq e^\epsilon\text{Pr}[\m(\e(\sigma))\in\s]\quad\forall \s\subseteq \o_M
\end{equation*}
where $\text{Pr}[\m(\e(\rho))\in\s]=\sum_{k\in\s}\tr(M_{k}\e(\rho))$.
\end{definition}
The provided definition elaborates on classical local differential privacy for randomized mechanisms (see Fig.~\ref{fig:framework}). It builds upon the same idea as local differential privacy, asserting that despite an adversary having access to an individual's personal responses, they should not be able to discern significant personal data due to the consistent behavior of the randomized mechanism across any pair of input states. This concept is extended to the quantum realm by introducing a parameter $\epsilon$ that ensures the similarity between the outcome probabilities $\sum_{k\in\s}\tr(M_{k}\e(\rho)) \text{ and }\sum_{k\in\s}\tr(M_{k}\e(\sigma))$
of quantum measurement $\m$ for any outcome subset $\s\subseteq\o_{\m}$ to manage the leakage of quantum state privacy under quantum mechanism $\e$. Later on, we will observe that this definition of QLDP aligns with the one presented in~\cite{hirche2023quantum} from a quantum information standpoint.

To draw a comparison between quantum and classical local differential privacy, we can utilize quantum measurement $\m$ as a (classical) randomized mechanism, as explained in Section~\ref{subsec:measurement}. Let us first revisit classical local differential privacy (LDP).

\begin{definition}\label{Def:LDP}
A randomized mechanism (mapping) $f$ is  $\epsilon$-\emph{locally differentially private (LDP)} if, for any states $x,y\in \text{domain}(f)$, the following condition is met:
\begin{equation*} \text{Pr}[f(x)\in\s]\leq e^\epsilon \text{Pr}[f(x)\in \s]\quad\forall \s\subseteq \text{image}(f)\end{equation*}
where $\text{domain}(f)$ and $\text{image}(f)$ denote the domain and image of the mapping $f(\cdot)$, respectively.
\end{definition}

For any quantum mechanism $\e$, we can derive a randomized mechanism $\m\circ\e$, which represents the function composition of $\e$ and $\m$. Subsequently, we can redefine QLDP of $\e$ by considering the LDP of $\m\circ\e$.

\begin{lemma}\label{lem:QLDP_LDP}
A quantum mechanism $\e$ is said to be $\epsilon$-QLDP if and only if the randomized mechanism $\m\circ\e$ exhibits $\epsilon$-LDP for any quantum measurement $\m$.
\end{lemma}

 Based on the aforementioned outcome, it is evident that in order to ensure an $\epsilon$-QLDP assurance for a quantum system, a quantum mechanism must ensure $\epsilon$-LDP for all quantum measurements conducted on the system simultaneously. This presents a significant challenge in safeguarding privacy against potential attackers, given the infinite number of quantum measurements possible within the system's Hilbert space $\h$. To overcome this difficulty, we apply eigenvalue analysis on the matrices describing the evolution of quantum mechanisms to study the essential quantitative properties of QLDP in the following.

\subsection{Evaluating QLDP}
In this section, we discuss the evaluation of the parameter $\epsilon$ for QLDP. Initially, we outline a method to calculate the classical LDP value of a randomized mechanism $f=\m\circ\e$ involving a quantum mechanism $\e$ and measurement $\m$. By utilizing this approach, we can determine the parameter $\epsilon$ of QLDP using Lemma~\ref{lem:QLDP_LDP}.

Before proceeding, it is necessary to introduce the Heisenberg picture, which tracks the evolution of quantum measurements, as opposed to the Schrödinger picture that monitors the evolution of quantum states as shown previously in Eq.(\ref{Eq:evolution}). To illustrate this, when considering a quantum measurement $\m$ with matrices $\{M_k\}_{k\in\o}$ and a quantum mechanism denoted by $\e$ with Kraus matrices $\{E_j\}$, given a quantum state $\rho$, the following calculations of the probability $\text{Pr}[\m(\e(\rho))=k]=\tr(M_k\e(\rho))$ of observing measurement outcome $k$ can be made:
\begin{equation*}
    \begin{aligned}
\tr(M_k\sum_jE_j\rho E_j^\dagger)=\tr(\sum_j E_j^\dagger M_jE_k \rho )=\tr(\e^\dagger(M_k) \rho ).
    \end{aligned}
\end{equation*}
In this context, the second equation is derived from the commutative nature of the trace operation, specifically, $\tr(AB)=\tr(BA)$, where $\e^\dagger(\cdot)=\sum_{j}E_j^\dagger \cdot E_j$ represents the dual (adjoint) mapping of $\e$ aimed at describing the evolution of quantum measurement $\m$.

With the Heisenberg picture and the dual mapping $\e$, we are able to analyze the LDP of randomized mechanism $\mathcal{M} \circ \e$.

\begin{theorem}\label{Thm:LDP}
    Suppose $\e$ represents a quantum mechanism and $\m$ represents a quantum measurement with matrices $\{M_k\}_{k\in\o_{\m}}$. The randomized mechanism $\m\circ\e$ is $\epsilon$-LDP if and only if 
    \[0 \geq \max_{\s\subseteq \o_{\m}}\lambda_{\max}[\e^\dagger(M_{\s})]-e^\epsilon\lambda_{\min}[\e^\dagger(M_{\s})].\]
    Here, $M_{\s}=\sum_{k\in\s}M_k$, and $\lambda_{\max}(M)$ and $\lambda_{\min}(M)$ refer to the maximum and minimum eigenvalues of the positive semi-definite matrix $M$, respectively. 
\end{theorem}

Therefore, the (optimal) LDP value $\epsilon^*$ (minimizing $\epsilon$) provided by the randomized mechanism $\m\circ\e$ can be determined as follows:
\begin{equation}\label{Eq:compute_LDP}
    \epsilon^*(\m\circ\e)=\max_{\s\subseteq \o_{\m}}\ln\frac{\lambda_{\max}[\e^\dagger(\sum_{k\in\s}M_k)]}{\lambda_{\min}[\e^\dagger(\sum_{k\in\s}M_k)]}.
\end{equation}
In cases where $\lambda_{\min}[\e^\dagger(\sum_{k\in\s}M_k)]=0$ for certain $\s\subseteq\o_{\m}$, the mechanism $\m\circ\e$ is considered to be $\infty$-LDP, indicating that the LDP leakage is unbounded.

Here, $\epsilon^*(\cdot)$ is considered a function determining the (optimal) LDP value for randomized mechanisms, which can also be extended to QLDP for quantum mechanisms. By utilizing Lemma~\ref{lem:QLDP_LDP}, we can express \[\epsilon^*(\e)=\max_{\m}\epsilon^*(\m\circ\e).\]

Calculation of $\epsilon^*(\m\circ\e)$ is essential for all quantum measurements $\m$ as indicated by the equation provided. It is worth noting that a quantum measurement can be defined by approximately $\bigO{N^4}$ parameters, where $N=2^n$ denotes the size of the state space in an $n$-qubit quantum system $\h$. This arises from a quantum measurement consisting of $N^2$ $N$-by-$N$ matrices at most~\cite[Theorem 2.1]{wolf2012quantum} and each matrix containing $N^2$ parameters. In the following, the objective is to gradually decrease the parameter count from $\bigO{N^4}$ to $\bigO{N}$ to enhance the efficiency of QLDP evaluation in quantum mechanisms.

In order to accomplish this, it can be observed that when $S^*$ denotes the optimal subset where $\epsilon^*(\m\circ \e)$ is achieved in Eq.~(\ref{Eq:compute_LDP}), we can establish a binary (outcome) quantum measurement $\m_{\s^*}=\setnocond{M_{\s^*}, I-M_{\s^*}}$. Thus, $\m\circ\e$ obeys $\epsilon$-LDP if and only if $\m_{\s^*}\circ\e$ complies with $\epsilon$-LDP. It is essential to recognize that $M_{\s^*}$ is a matrix falling within the set $\{M:0\leq M\leq I\}$. For any positive semi-definite matrices $M_1$ and $M_2$, $M_1\leq M_2$ if and only if $\bra{\psi}M_1\ket{\psi}\leq \bra{\psi}M_2\ket{\psi}$ holds for all pure quantum state $\ket{\psi}$. Moreover, any $0\leq M\leq I$ can function as a matrix contributing to a quantum measurement (for instance, a binary one $\m_{M}$ with matrices $\{M, I-M\}$). Such a matrix $0\leq M\leq I$ is referred to as a \emph{measurement matrix}. Consequently, we obtain 
\[\epsilon^*(\e)=\max_{0\leq M\leq I}\epsilon^*(\m_{M}\circ\e).\]

\begin{remark} In the earlier definition of quantum (local) differential privacy, the focus was primarily on defining them for all measurement matrices $\{M: 0\leq M\leq I\}$~\cite{hirche2023quantum,nuradha2023quantum} without specifying the actual quantum measurements. By considering the characterization of QLDP value mentioned above, it becomes evident that our definition of QLDP in Definition~\ref{Def:QLDP} is consistent with those previously established. Consequently, a post-process theorem can be directly deduced from prior research on quantum differential privacy in the next section. Nonetheless, the other findings regarding QLDP in this study cannot be derived using the same approach since earlier studies rely on quantum divergences, which are effective for analyzing specific pairs of quantum states to investigate fundamental properties of quantum differential privacy. As highlighted in the related work section, analyzing all quantum states simultaneously poses a challenge when exploring optimal utility-privacy mechanisms within the QLDP framework. To address this issue, we will introduce an eigenvalue-based representation of QLDP value across all pure states in the following, which will prove instrumental in identifying the optimal quantum mechanisms.
\end{remark}

By combining these insights with Theorem~\ref{Thm:LDP}, we can derive the following description of QLDP for the quantum mechanism $\e$.

\begin{lemma}\label{Lem:QLDP_single}
    A quantum mechanism $\e$ is  $\epsilon$-QLDP if and only if, for any measurement matrix $0\leq  M\leq I$, the following inequality holds:
    \[0 \geq \max_{0\leq M\leq I}\lambda_{\max}[\e^\dagger(M)]-e^\epsilon\lambda_{\min}[\e^\dagger(M)].\]
\end{lemma}
Following the above result, our attention turns to managing an $N$-by-$N$ matrix $M$, which can be expressed through $N^2$ parameters based on the aforementioned outcome. To decrease the number of parameters even further, we can utilize the eigendecomposition of the measurement matrix $M=\sum_{k}\lambda_k\ketbra{\psi_k}{\psi_k}$ with eigenvalue $\lambda_k\geq 0$, as illustrated in Eq.(\ref{Eq:ensemble}) by noting that the normalized $\frac{M}{\tr(M)}$ is a quantum state.

 \begin{theorem}\label{Thm:QLDP}
    A quantum mechanism $\e$ acting on a Hilbert space $\h$ is $\epsilon$-QLDP if and only if  $\epsilon\geq \epsilon^*(\e)$, where the optimal QLDP value $\epsilon^*(\e)$ of $\e$ is determined by the formula: 
    \[\epsilon^*(\e)=\max_{\ket{\psi}\in\h}\ln\frac{\lambda_{\max}[\e^\dagger(\psi)]}{\lambda_{\min}[\e^\dagger(\psi)]}. \]
    If $\lambda_{\min}[\e^\dagger(\psi)]=0$ for some $\psi$, then we say $\e$ is $\infty$-QLDP, denoted as $\epsilon^*(\e)=\infty$, indicating unbounded QLDP leakage. 
\end{theorem}
Here, it is worth recalling that $\psi=\ketbra{\psi}{\psi}$ represents the mixed state form corresponding to the pure state $\ket{\psi}$. Based on the outcome mentioned above, we can calculate the QLDP value by addressing an optimization problem involving the variables represented by $\ket{\psi}$, which encompass around $N$ parameters for $(N=2^n)$-dimensional quantum system $\h$. This approach aims to minimize the parameter count to $\bigO{N}$ as previously discussed for enhancing efficiency.

By utilizing Theorems~\ref{Thm:LDP} and~\ref{Thm:QLDP}, we can calculate the LDP and QLDP values for randomized mechanisms and quantum mechanisms, correspondingly. Examining examples of quantum noise mechanisms can help illustrate the distinction between these two quantities.
 \begin{example}
Consider two measurements $\m_0$ and $\m_1$ denoted by measurement matrices $\{\ketbra{0}{0},\ketbra{1}{1}\}$ and $\{\frac{1}{2}I,\frac{1}{2}I\}$, respectively. When subjected to depolarizing noise mechanism $\n_{\dep}$ with probability $p$ as defined in Example~\ref{Exa:noise}, the QLDP value is given as $\epsilon^*(\n_{\dep})=\ln\frac{1+p}{1-p}$, while the  LDP value is expressed as 
\[\epsilon^*(\m_0\circ\n_{\dep})=\ln\frac{1+p}{1-p} \text{ and }\epsilon^*(\m_1\circ\n_{\dep})=0.\]

It is evident that the QLDP value $\epsilon^*(\n_{\dep})$ can be reached using the LDP value $\epsilon^*(\m_{0}\circ\n_{\dep})$. This suggests that by employing quantum measurement $\m_{0}$, an attacker can attain the maximum QLDP leakage offered by the quantum noise mechanism $\n_{\dep}$. On the other hand, utilizing quantum measurement $\m_1$ alone does not provide any information on quantum state privacy, as the corresponding LDP value is 0.
\end{example}
In the simplest scenario of quantum algorithms and channels, a quantum mechanism $\e=\u$ can be represented by a single Kraus (unitary) matrix U ($UU^\dagger=I$), where $\u(\rho)=U\rho U^\dagger$ for all $\rho\in\dh$. In such instances, the QLDP of $\u$ can be directly obtained from Theorem~\ref{Thm:QLDP} in the worst-case scenario. 
\begin{corollary}\label{Coro:LDP_post}
    Any unitary quantum mechanism $\u$ is  $\infty$-QLDP.
\end{corollary}

As stated in Corollary~\ref{Coro:LDP_post}, unitary quantum systems are highly vulnerable to attacks in the QLDP scenario. Therefore, it is imperative to establish a defense mechanism to safeguard the privacy of quantum states. In classical local differential privacy, one can provide privacy guarantee by incorporating noise into the classical data before collecting. Similarly, in the realm of quantum systems, this can be achieved through the application of a quantum post-processing theorem, as elaborated in the subsequent section.

\begin{remark}
    While we introduce the computation of the QLDP value $\epsilon^*$ as an optimization problem with $\bigO{N}$ variables in Theorem~\ref{Thm:QLDP}, we do not present a solution algorithm or conduct a complexity analysis. Our focus is on demonstrating the practical application of the QLDP framework by identifying optimal utility-privacy quantum mechanisms. The method to solve this problem for an arbitrary $n$-qubit system remains unclear. However, we can subsequently determine a closed form of $\epsilon^*$ for systems with small dimensions using symbolic execution, exemplified by the 1-qubit noise mechanism in Theorem~\ref{Exa:noise_cost}. Furthermore, the structure of the optimization problem outlined in Theorem~\ref{Thm:QLDP} serves as a useful tool to showcase the effectiveness of depolarizing noise in maximizing utility while adhering to the $\epsilon^*$-QLDP constraint, as discussed in Section~\ref{Sec:experiment}. Developing an efficient algorithm to solve this optimization problem is an important and challenging endeavor, building upon the foundation of Theorem~\ref{Thm:QLDP} to evaluate quantum mechanisms' QLDP value $\epsilon^*$. We aim to delve into this task in an upcoming project, including investigating the problem's decidability and, if determinable, classifying it among optimization problems like semi-definite programming and linear programming commonly employed in quantum information research.
\end{remark}


 \subsection{Post-processing Guarantee and Effective Noise Mechanisms}
In this section, we first present a post-processing theorem from the studies~\cite{hirche2023quantum,nuradha2023quantum} for QLDP to demonstrate that incorporating quantum noise mechanisms is an effective strategy for safeguarding the privacy of quantum states. Then, we outline a method to pinpoint efficient quantum noise mechanisms that yield a measurable QLDP value.
\begin{proposition}[\cite{hirche2023quantum,nuradha2023quantum}]\label{Thm:post-process}
    If a quantum mechanism  $\e$ is $\epsilon$-QLDP, then the (functionally) composed quantum mechanism $\n\circ\e$ is still $\epsilon$-QLDP for any quantum mechanism $\n$. 
\end{proposition} 
According to this post-processing result, we can utilize a quantum noise mechanism to avoid privacy breaches. The challenge now lies in selecting the appropriate mechanism, which should offer a finite QLDP value at the very least. To address this, it is essential to initially assess whether $\epsilon^*(\e)<\infty$ for a specified quantum mechanism $\e$.

\begin{theorem}\label{Lem:QLDP_checking}
    Consider a quantum mechanism $\e$ with Kraus matrices $\{E_k\}_{k=0}^{K-1}$  on a Hilbert space $\h$. The QLDP value $\epsilon^*(\e)< \infty$ holds if and only if 
    one of the following equivalent conditions is met:
    \begin{enumerate}
    
   \item $\e^\dagger\otimes \i(\ketbra{\Omega}{\Omega})>0$, where $\ket{\Omega}=\sum_{j}\ketbra{j,j}{j,j}$ is the unnormalized maximal entangled state over composed quantum system $\h\otimes\h$ and $\{\ket{j}\}_{j}$ is a mutually orthogonal basis of $\h$;
   \item The linear span of $\{E_k^\dagger\}$ encompasses the entire matrix space on $\h$, indicating that any matrix on $\h$ can be linearly represented by $\{E_k^\dagger\}$.
    \end{enumerate}
Checking either of the two conditions mentioned above has a time complexity of $\bigO{N^6K}$, where $N$ represents the dimension of $\h$ and $K$ stands for the number of Kraus matrices of $\e$.
\end{theorem}
Based on the second condition, we can infer from linear algebra principles that a minimum of $N^2$ Kraus matrices is required to cover the entire matrix space on a Hilbert space $\mathcal{H}$ of dimension $N$. Consequently, the number $K$ of Kraus matrices $\{E_k\}_{k=0}^{K-1}$ in a quantum system $\mathcal{E}$ effectively indicates that $\epsilon^*(\mathcal{E})=\infty$.
\begin{corollary}\label{cor:QLDP_effectiveness}
    Consider a quantum mechanism $\e$ with Kraus matrices $\{E_k\}_{k=1}^K$  on a Hilbert space $\h$ with dimension $N$. If $K<N^2$, then $\epsilon^*(\e)=\infty$.
\end{corollary}
By the corollary, it is evident that some quantum noise mechanisms outlined in Example~\ref{Exa:noise} do not yield a quantifiable QLDP value. In contrast, for the rest quantum noises, we can determine the QLDP value $\epsilon ^*$ by solving the optimization problem shown in Theorem~\ref{Thm:QLDP}.
\begin{theorem}[QLDP Values]\label{Exa:noise_cost}
    For the standard 1-qubit noises presented in Example~\ref{Exa:noise}, the (optimal) QLDP values provided by them are as follows. 
    \begin{enumerate}
        \item For the Pauli-type noises with probability $0\leq p\leq 1$, we have
        \[\epsilon^*(\n_\text{XF})=\epsilon^*(\n_\text{YF})=\epsilon^*(\n_\text{ZF})=\infty\] and 
        \begin{equation*}
        \epsilon^*(\n_{\dep})=\left\{\begin{aligned}
            &\ln\frac{1+p}{1-p} &\text{ if} \quad 0\leq p<1 \\
            &\infty  &\text{ if} \quad p=1 
        \end{aligned}\right. .
        \end{equation*}
        \item For the damping-type noises with parameters $0\leq r,q \leq 1$, we have 
        \[ \epsilon^*(\n_\text{PD})=\epsilon^*(\n_\text{AD})=\infty \]
        and furthermore, for $q=0,1$ or $r=0$, $\epsilon^*(\n_{\gad})=\infty.$ Otherwise, 
        \[\epsilon^*(\n_{\gad})=\ln(\frac{\sqrt{1-rh^2}+\sqrt{1-r}}{\sqrt{1-rh^2}-\sqrt{1-r}})\]
        where $h=|1-2q|$.
        
    \end{enumerate}    
\end{theorem}


Moreover, when considering these efficient quantum noise mechanisms such as $\n_{\dep}$ and $\n_{\gad}$, the key issue is determining their relative effectiveness. In addressing this concern, utility serves as a crucial criterion for selecting the most suitable noise mechanism within the classical realm. Therefore, the subsequent section will focus on identifying the most optimal mechanism in terms of utility to effectively manage the trade-off between QLDP and utility.

\begin{remark}
    
 By Theorem~\ref{Exa:noise_cost}, many quantum noise mechanisms do not yield a finite QLDP value necessary for safeguarding the privacy of quantum states. This limitation may result from the stringent confinement of QLDP as defined in Definition~\ref{Def:QLDP}, which relies solely on the parameter $\epsilon$. To address this issue, we may consider loosening the requirement of achieving $\epsilon$-QLDP by introducing an additional parameter $\delta$ to manage privacy breaches. This approach follows the same idea as the classical case. Consequently, we introduce the following revised definition:
\begin{definition}
A quantum mechanism $\e$ acting on a Hilbert space $\h$ is  $(\epsilon,\delta)$-\emph{quantum locally differentially private (QLDP)} if, for any quantum measurement $\m=\{M_{k}: k\in\o_{\m}\}$ and any quantum states $\rho,\sigma\in\dh$, the following condition holds:
\begin{equation*} \sum_{k\in\s}\tr(M_{k}\e(\rho))\leq e^\epsilon\sum_{k\in\s}\tr(M_{k}\e(\sigma))+\delta\quad\forall \s\subseteq \o_M.\end{equation*}
\end{definition}

Despite the relaxation in the QLDP definition, it is observed that $\n_\text{XF},\n_\text{ZF},\n_\text{PD}$, and $\n_\text{AD}$ in Example~\ref{Exa:noise} can only ensure $(\epsilon,1)$-QLDP for any $\epsilon$, which is unsurprising given that $\delta\leq 1$. Consequently, this relaxed version of QLDP may not be the most suitable for quantum computing, necessitating the exploration of alternative methods to incorporate these noises effectively as a defensive mechanism. We left this as a future work. 
\end{remark}
\section{Optimal Mechanisms for Utilities}\label{sec:optimal}
In this section, we demonstrate that the depolarizing noise mechanism is the best choice for ensuring an $\epsilon$-QLDP guarantee while maximizing fidelity and trace distance utilities across all dimensions of quantum state space compared to other unital quantum mechanisms.
\subsection{Choices of Utilities}
Within our threat model framework in Section~\ref{Sec:threat_model}, the key emphasis lies in protecting quantum states. The incorporation of quantum noise mechanisms is crucial to achieving this objective but can potentially compromise the integrity of these quantum states. Consequently, this compromises the precision and utility of quantum states for specific analytical purposes. Therefore, it becomes essential to assess the effectiveness of a quantum noise mechanism in retaining information pertaining to quantum states. Traditionally, the evaluation of such utilities is often conducted through fidelity and anti-trace distance within the quantum realm.

The two utilities of quantum noise mechanisms are based on fidelity and trace distance of a pair of quantum states. The fidelity $F(\rho,\sigma)$ and trace distance $T(\rho,\sigma)$ for two quantum states $\rho$ and $\sigma$ are defined as follows:
\[F(\rho,\sigma)\coloneqq  [\text{tr}(\sqrt{\sqrt{\rho}\sigma\sqrt{\rho}})]^2 \quad \text{and} \quad T(\rho,\sigma) \coloneqq\frac{1}{2}\text{tr}(|\rho-\sigma|).\]
For any Hermitian matrix $H$\footnote{A matrix $H$ is considered Hermitian when $H^\dagger=H$. Consequently, all eigenvalues of $H$ are real numbers.} (e.g., $\rho$ and $\rho-\sigma$) with eigendecomposition $\sum_{k}\lambda_k\ketbra{\psi_k}{\psi_k}$, the square root of $H$ is defined as $\sqrt{H}=\sum_{k}\sqrt{\lambda_k}\ketbra{\psi_k}{\psi_k}$ and the absolute value of $H$ is denoted as $|H|=\sum_{k}|\lambda_k|\ketbra{\psi_k}{\psi_k}$.
The fidelity of two states quantifies their similarity, while trace distance measures their dissimilarity. To see this, for example,  $F(\rho,\sigma)=1$ ($T(\rho,\sigma)=0$) if and only if $\rho=\sigma$, indicating that $\rho$ and $\sigma$ are indistinguishable. On the other hand, $F(\rho,\sigma)=0$ ($T(\rho,\sigma)=1$) if and only if $\rho$ is orthogonal to $\sigma$, implying that $\rho$ and $\sigma$ are entirely distinguishable~\cite{nielsen2010quantum}.

We can now establish the concepts of fidelity and anti-trace distance utilities for quantum mechanisms with respect to all quantum states in the worst-case scenario~\cite[Section 9.3]{nielsen2010quantum}.

\begin{definition}
The fidelity utility of a quantum mechanism $\n$ on a Hilbert space $\h$ is defined as
\[F(\n) \coloneqq \min_{\rho \in \dh} F(\n(\rho), \rho).\]
\end{definition}
By definition, the fidelity utility $F(\n)$ is determined by the minimal information (utility) preserved by $\n$ on all quantum states based on the fidelity measure. 
\begin{definition}
The anti-trace distance utility of a quantum mechanism $\n$ on a Hilbert space $\h$ is defined as:
\[\hat{T}(\n) \coloneqq 1 - T(\n) \quad
\text{where} \quad
T(\n) \coloneqq \max_{\rho \in \dh} \|\n(\rho) - \rho\|.\]
\end{definition}
Defined as $T(\n)$, this metric quantifies the maximal noisy effects caused by $\n$ on all quantum states using the trace distance. Consequently, $\hat{T}(\n)$ evaluates the minimal information (utility) preserved by  $\n$ on all quantum states.

To estimate the two utilities, by the joint concavity of fidelity and trace distance~\cite[Section 9.3]{nielsen2010quantum}, we only need to focus on pure quantum states as follows.
    \begin{equation*}
        \begin{aligned}
           & F(\n)=\min_{\ket{\psi}\in\h}F(\n(\psi), \psi)=\min_{\ket{\psi}\in\h}\bra{\psi}\n(\psi)\ket{\psi}\\
           & \hat{T}(\n)=1-\max_{\ket{\psi}\in\h}T(\n(\psi), \psi).
        \end{aligned}
    \end{equation*}
With the above equations, we can analytically compute the fidelity and anti-trace distance utilities for the quantum noise mechanisms introduced in Example~\ref{Exa:noise}. 
\begin{theorem}[Utilities for Quantum Noise Mechanisms]\label{Exa:utility}
    For the standard 1-qubit noises presented in Example~\ref{Exa:noise}, the utilities provided by them are as follows. 
    \begin{enumerate}
        \item For the Pauli-type noises with noiseless probability $0\leq p\leq 1$, we have
    \begin{equation*}
    \begin{aligned}
        &F(\n_\text{UF})=\hat{T}(\n_\text{UF})=p \quad\text{ for }\quad U\in\{X, Y, Z\} \\
        &F(\n_{\dep})=\hat{T}(\n_\dep)=\frac{1+p}{2}.\\
    \end{aligned}
\end{equation*}

        \item For the damping-type noises with parameters $0\leq r,q \leq 1$, we have 
        \begin{equation*}
    \begin{aligned}
        &F(\n_\text{PD})=\hat{T}(\n_\text{PD})=\frac{1+\sqrt{1-r}}{2}\\
        &F(\n_{AD})=\hat{T}(\n_{AD})=1-r\\
        &F(\n_{\gad})=\hat{T}(\n_{\gad})=1-\frac{1}{2}(h+1)r=\left\{
    \begin{aligned}
        &1-qr &\text{ if } q\geq \frac{1}{2}\\
        &1-(1-q)r &\text{ if } q< \frac{1}{2}
    \end{aligned}\right. 
    \end{aligned}
\end{equation*}  
    \end{enumerate}    
where $h=|1-2q|$.
\end{theorem}
As we can see, parameters $p, q, $ and $r$ can be utilized to regulate the utilities offered by quantum noise mechanisms. When considering the QLDP value illustrated in Theorem~\ref{Exa:noise_cost}, determining the most suitable option becomes an intriguing challenge.

By utilizing Theorems~\ref{Exa:noise_cost} and~\ref{Exa:utility}, we can establish the quantitative connection between utility and privacy concerning 1-qubit depolarizing and generalized amplitude damping noise mechanisms with $0\leq p<1$, $0<q<1$ and $0< r\leq 1$. It is important to note that other noise mechanisms in Example~\ref{Exa:noise} do not yield significant QLDP values in this context, as shown in Theorem~\ref{Exa:noise_cost}.
\begin{equation}\label{Eq:dep_utility}
    \begin{aligned}
        &F(\n_{\dep})=\hat{T}(\n_{\dep})=\frac{e^\epsilon}{e^\epsilon+1} \quad\text{ for }\quad \epsilon=\epsilon^*(\n_{\dep})
    \end{aligned}
\end{equation}
\begin{equation}\label{Eq:gad_utility}  
F(\n_\gad)=\hat{T}(\n_\gad)=1-\frac{1}{2}(h+1)\frac{\beta^2-1}{\beta^2-h^2}
\end{equation}
    where  $h=|1-2q|$, $\beta=\frac{e^\epsilon+1}{e^\epsilon-1}$ and $\epsilon=\epsilon^*(\n_{\gad})$.
    
The forms above illustrate the balance between the utilities and QLDP value of the two quantum noises. Additionally, we can compare these quantum noises within the QLDP framework. Further elaboration on this comparison will be presented in the subsequent analytical analysis and numerical experiments in Section~\ref{Sec:experiment}.

\begin{remark}
    
Although Theorem~\ref{Exa:utility} demonstrates that for any standard 1-qubit quantum noise mechanism introduced in Example~\ref{Exa:noise}, the fidelity utility is equal to the anti-trace distance utility, it remains uncertain whether this equality holds true for all quantum mechanisms. Instead, a general inequality holds for any quantum mechanism $\n$:
\[\hat{T}(\n)\leq  F(\n).\]

To elaborate, the relationship between trace distance and fidelity states that for any quantum states $\rho$ and $\sigma$, the trace distance provides upper and lower bounds on the fidelity, as indicated by the Fuchs–van de Graaf inequalities~\cite{fuchs1999cryptographic}:
\begin{equation}\label{Eq:fidelity_trace}
{\displaystyle 1-{\sqrt {F(\rho ,\sigma )}}\leq T(\rho ,\sigma )\leq {\sqrt {1-F(\rho ,\sigma )}}.\,}
\end{equation}
When at least one of the states is a pure state $\psi$, the lower bound can be further refined to:
    \[1-F(\rho,\psi)\leq T(\rho,\psi).\]
    Consequently, through the arbitrariness of $\rho$, we deduce
    \[1-T(\n(\psi),\psi)\leq F(\n(\psi),\psi).\]
By optimizing all pure states independently on both sides, we derive
\[\min_{\psi}[1-T(\n(\psi),\psi)]\leq \min_{\psi}F(\n(\psi),\psi).\]
This leads to the conclusion $\hat{T}(\n)\leq  F(\n).$
It is important to note that the trace distance is often more computationally feasible to evaluate or constrain compared to fidelity, making these relationships particularly valuable. 

In this paper, we examine the balance between the utility and privacy concerns of quantum noise mechanisms. We do not currently address the equivalence of these benefits and consider it a topic for future research.
\end{remark} 

\subsection{Optimal Mechanisms}
This section is dedicated to the identification of optimal unital quantum mechanisms that maximize information theoretic utilities while upholding a constant QLDP value. Specifically, we focus on fidelity and trace distance utilities within an $n$-qubit quantum system $\h$ for practical implementation.

We address two main problems formally:
\begin{problem}[Fidelity utility optimization]\label{prob:fidelity}
In the case of fidelity utility, the objective is to maximize fidelity utility among all $\epsilon$-QLDP quantum mechanisms, given by:
\begin{align*}
    \max_{\n}\quad & F(\n)\\
\text{subject to} \quad&\n\in\nn_{\epsilon}.
\end{align*}
Here, $\nn_{\epsilon}=\{\n:\epsilon^*(\n)=\epsilon \text{ and } \n(I)=I\}$ represents the set of unital quantum mechanisms with a QLDP value of $\epsilon$.
\end{problem}

\begin{problem}[Anti-trace distance utility optimization]\label{prob:trace}
In the case of trace distance utility, the goal is to maximize the anti-trace distance utility $\hat{T}$ among all $\epsilon$-QLDP quantum mechanisms, given by:
\begin{align*}
    \max_{\n}\quad & \hat{T}(\n)\\
\text{subject to} \quad&\n\in\nn_{\epsilon}.
\end{align*}
\end{problem}
To address the two optimization issues mentioned above, we initially establish the optimal bounds for both utilities by utilizing the QLDP value as depicted below.

\begin{theorem}\label{Thm:tradeoff} For any unital quantum mechanism $\n$ with $\epsilon^*(\n)<\infty$, we have 
\begin{equation*}
    \begin{aligned}    
         F(\n)\leq \frac{e^{\epsilon}}{e^{\epsilon}+2^n-1}\qquad
        \hat{T}(\n)\leq \frac{e^{\epsilon}}{e^{\epsilon}+2^n-1}.
    \end{aligned}
\end{equation*}
 Here $\epsilon=\epsilon^*(\n)$ and either of these equalities is obtained if and only if $\n\in\nn^*$, where the set of optimal quantum mechanisms $\nn^*$ is defined as
   \[\{\n_{\eta}:\n^\dagger_{\eta}(\rho)=\eta\rho+\frac{1-\eta}{2^n-1}(I-{\rho}) \text{ and } 1>\eta\geq \frac{1}{2^n}\}.\]
\end{theorem}
Hence, based on the aforementioned outcome, our primary focus should be on the set $\nn^*$ to maximize utilities. Let's examine the scenario of a 1-qubit system, where $n=1$. In this instance, $\nn^*$ is represented by the set 
\[\{\n_{\eta}: \n^\dagger_{\eta}(\rho)=(2\eta-1)\rho+(1-\eta)I\text{ and } 1>\eta\geq \frac{1}{2}\}.\]
If we let $p=2\eta-1$, we can express $\nn^*$ as 
\[\{\n: \n^\dagger(\rho)=p\rho+\frac{1-p}{2}I\text{ and }1>p\geq 0\}.\]
Therefore, in a 1-qubit system, $\nn^*$ is comprised of depolarizing quantum noise processes with a probability $p<1$, where depolarizing noise $\n_{\dep}$ is its own adjoint. Extending this concept to a general $n$-qubit quantum system, the parameter $p$ defined as $p=\eta-\frac{1-\eta}{2^n-1}$ results in:
\[\nn^*=\{\n:\n^\dagger(\rho)=p\rho+(1-p)\frac{I}{2^n} \text{ and } 0\leq p <1\}.\]
Under this definition, all quantum processes within $\nn^*$ are essentially $n$-qubit depolarizing noises with a probability less than 1. The formal representation of a depolarizing noise, denoted as $\n_{\dep}^n$, acting with a probability of $p$ on an $n$-qubit quantum state space $\h$~\cite{nielsen2010quantum}, is given by 
\[\n_{\dep}^n(\rho)=p\rho+(1-p)\frac{I}{2^n} \qquad\forall \rho\in \dh.\]
Furthermore, the adjoint of $\n_{\dep}^n$ is itself.
Specifically, the depolarizing noise for a 1-qubit system is labeled as $\n_{\dep}^1$. Therefore, the optimal set $\nn^*$ in Theorem~\ref{Thm:tradeoff} encompasses all depolarizing noises with a probability $p<1$. It is worth noting that the adjustment of utilities and QLDP values of depolarizing noise can be achieved by manipulating the probability $p$.
\begin{lemma}\label{Lem:dep_utility_QLDP}
    For a depolarizing quantum noise $\n_{\dep}^n$ with probability $0\leq p <1$ on a $n$-qubit system, we have
    \begin{itemize}
    \item QLDP value is 
    \[\epsilon^*(\n_{\dep}^n)=\ln[1+2^n(\frac{1}{1-p}-1)].\]
    \item  Fidelity and anti-trace distance utilities both are 
    \[F(\n_{\dep}^n)=\hat{T}(\n_{\dep}^n)=\frac{(2^n-1)p+1}{2^n}.\]
    \end{itemize}
\end{lemma}
This result indicates that as the increment of $p$ occurs, the QLDP value and utilities also increase. This demonstrates a clear quantitative trade-off between the utilities and privacy of depolarizing noise mechanisms. By combining these values based on $p$, we can address the optimization problems related to fidelity and trace distance utility (Problems~\ref{prob:fidelity} and~\ref{prob:trace}) as follows.

\begin{theorem}\label{thm:solution}
    The optimal solutions for the fidelity and anti-trace distance utility optimization problems (Problems~\ref{prob:fidelity} and~\ref{prob:trace}) in an $n$-qubit system are given by
    \[F(\n^*)=\hat{T}(\n^*)=\frac{e^\epsilon}{e^\epsilon+2^n-1},\]
    where $\n^*$ represents the quantum mechanism leading to the optimal values. All feasible solutions contribute to a set denoted as $\nn^*_{\epsilon}$,
    \[\nn^*_{\epsilon}=\{\n_{\dep}^n:\n_{\dep}^n(\rho)=\frac{e^\epsilon-1}{e^\epsilon+2^n-1}\rho+\frac{I}{e^\epsilon+2^n-1}\}.\]
\end{theorem}

According to the above findings, the depolarizing noise mechanism emerges as the preferred option for balancing the trade-off between QLDP and either fidelity or trace distance utility.

Finally, it is evident that depolarizing noise serves as a quantum extension of classical randomized responses~\cite{warner1965randomized}. To illustrate this point, we first consider the depolarizing noise mechanism denoted as $\mathcal{N}_{\dep}^n$. According to Theorem~\ref{thm:solution}, the behavior of $\mathcal{N}^n_{\dep}$ with QLDP value $\epsilon$ can be expressed as follows:

\begin{equation}
    \begin{aligned}
        \mathcal{N}_{
        \dep}^n(\rho) = \frac{e^\epsilon}{e^\epsilon+2^n-1}\rho + \left(\frac{1}{e^\epsilon+2^n-1}\right)(I-\rho).
    \end{aligned}
\end{equation}

Next, we introduce a quantum measurement $\{\psi_k\}_{0\leq k\leq 2^n-1}$ comprising a collection of mutually orthogonal pure states $\ket{\psi_k}$. Considering the observation of quantum state $\psi_{k}$ given quantum state $\psi_{j}$ under depolarizing noise $\mathcal{N}_{\dep}^n$, the conditional probability can be formulated as:
\begin{equation*}
\text{Pr}(X=k|Y=j) = \text{Tr}(\psi_k\mathcal{N}_{\dep}^n(\psi_j)) = \left\{
    \begin{aligned}
        &\frac{e^\epsilon}{e^\epsilon+2^n-1} &\text{ if } k=j\\
        &\frac{1}{e^\epsilon+2^n-1} &\text{ if } k \neq j
    \end{aligned}\right . .
\end{equation*}
In this scenario, the depolarizing noise involves a straightforward randomization across the measurement outcome $\{0,1,\cdots,2^n-1\}$, where the actual data is disclosed with a probability of $\frac{e^\epsilon}{e^\epsilon+2^n-1}$. This mirrors the concept of classical randomized response as discussed in~\cite{kairouz2016extremal}. Therefore, depolarizing noise can be seen as an extension of randomized response. This extension provides a clearer insight into the effectiveness of safeguarding QLDP, given that the randomized response mechanism has been shown to optimize the KL divergence utility across all classical $\epsilon$-LDP mechanisms~\cite{kairouz2016extremal}.

\begin{remark}   
 In certain extreme scenarios, our objective is to minimize the QLDP value concerning a fixed anti-trace distance or fidelity utility. According to Theorem~\ref{Thm:tradeoff}, it holds for any unital quantum mechanism $\n$ that:
\begin{equation*}
    \begin{aligned}
        &\epsilon^*(\n)\geq\ln[(2^n-1)(\frac{1}{1-F(\n)}-1)]\\
        &\epsilon^*(\n)\geq\ln[(2^n-1)(\frac{1}{1-\hat{T}(\n)}-1)].
    \end{aligned}
\end{equation*}
Consequently, the minimum value of $\epsilon^*(\n)$ is only attained when $\n\in\nn^*$. Thus, depolarizing noise emerges as the optimal mechanism for minimizing the QLDP value across all quantum mechanisms with either a fixed fidelity or anti-trace distance utility.
\end{remark}

\section{Experimental Evaluation}\label{Sec:experiment}
In this section, we conduct numerical experiments on three aspects to demonstrate the efficacy of depolarizing noise in the QLDP framework.
\begin{enumerate}
\item \emph{Utility Optimization:} In comparison to different (non-unital) general amplitude damping noise mechanisms, we establish the optimization of fidelity (anti-trace distance) utility across varying QLDP values.
\item \emph{Trade-off:} By adjusting the noiseless probability $p$ to attain the targeted QLDP value and depolarizing utility, we can observe the trade-off between these two aspects.
\item \emph{Qubit Number:} As we scale up the number of qubits in quantum systems, we notice a significant exponential decline in fidelity utility for a specific QLDP value.
\end{enumerate}
It is important to mention that since the fidelity and anti-trace distance utilities of depolarizing channels are equal, we will focus solely on the fidelity utility in the upcoming experiments. 

Depolarizing noise is readily implementable on real quantum devices, with support from mainstream quantum computing cloud platforms such as IBM Qiskit and Google Cirq, further enhancing its practical applicability.
\begin{figure}[htp]
    \centering
    \includegraphics[width=\linewidth]{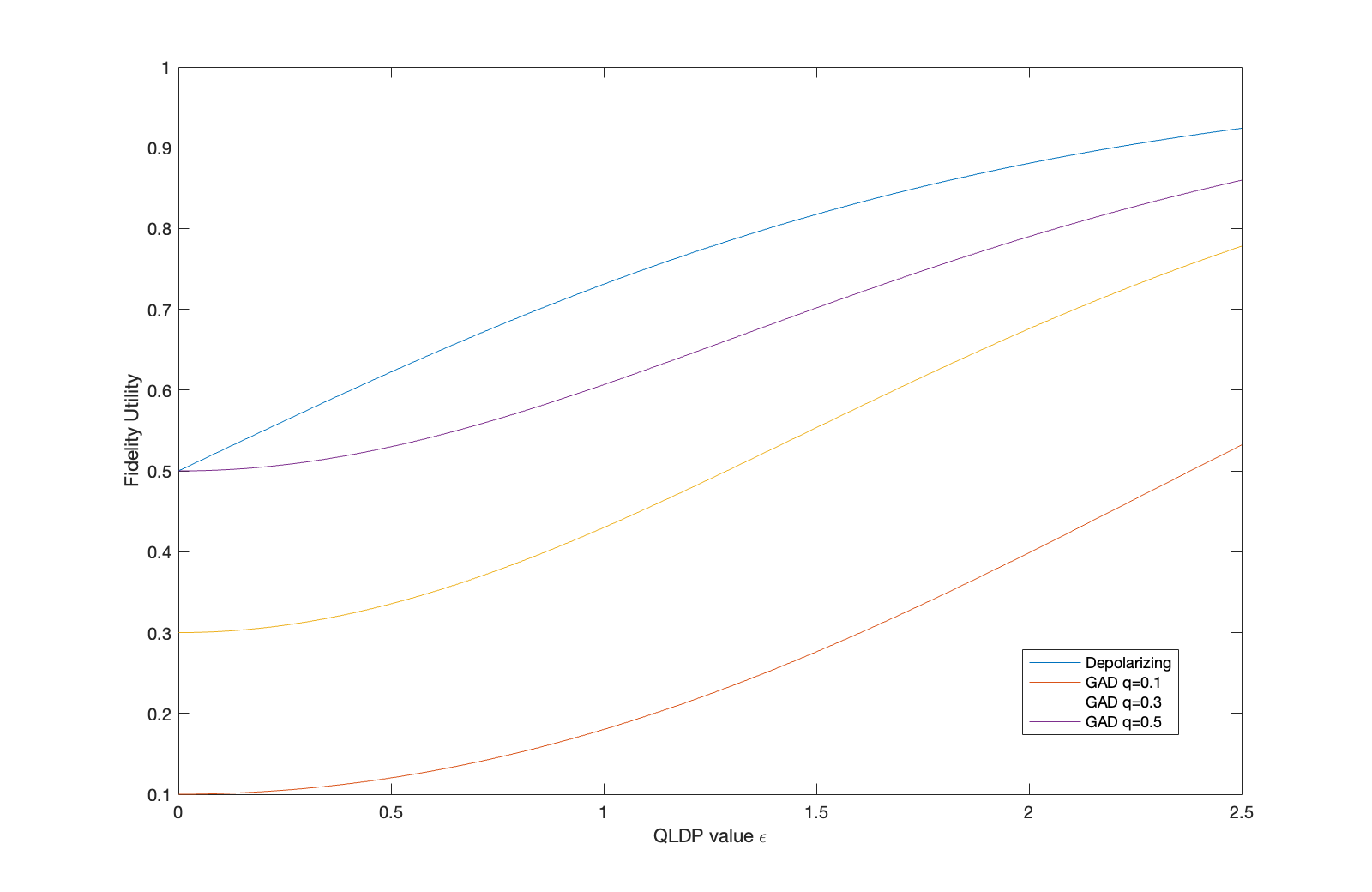}
    \caption{The compared fidelity utility of  depolarizing and generalization amplitude damping noise mechanisms with various parameter $q$.}
    \label{fig:utility_optimization}
\end{figure}
\subsection{Utility Optimization}
We have demonstrated the optimization of the fidelity and anti-trace distance utilities with depolarizing noise within unital quantum mechanisms in Theorem~\ref{thm:solution}. The next question to address is whether the effectiveness of the depolarizing noise mechanism surpasses that of non-unital quantum mechanisms, such as generalized amplitude damping noises, in this context. To explore this, we compare the fidelity utilities of 1-qubit depolarizing noise and generalized amplitude damping noise mechanisms by varying their QLDP values. By utilizing Eqs.(\ref{Eq:dep_utility}) and (\ref{Eq:gad_utility}), we are able to graph the fidelity utility function using the QLDP value as the variable for 1-qubit depolarizing noise and various generalized amplitude damping noise types characterized by parameter $q$. Refer to Fig.~\ref{fig:utility_optimization} for a visual comparison. It is evident that the utility derived from depolarizing noise surpasses that of any generalized amplitude damping noise at any given QLDP value. This observation underscores the effectiveness of depolarizing noise in safeguarding quantum state information.

\begin{figure}[htp]
    \centering
    \includegraphics[width=\linewidth]{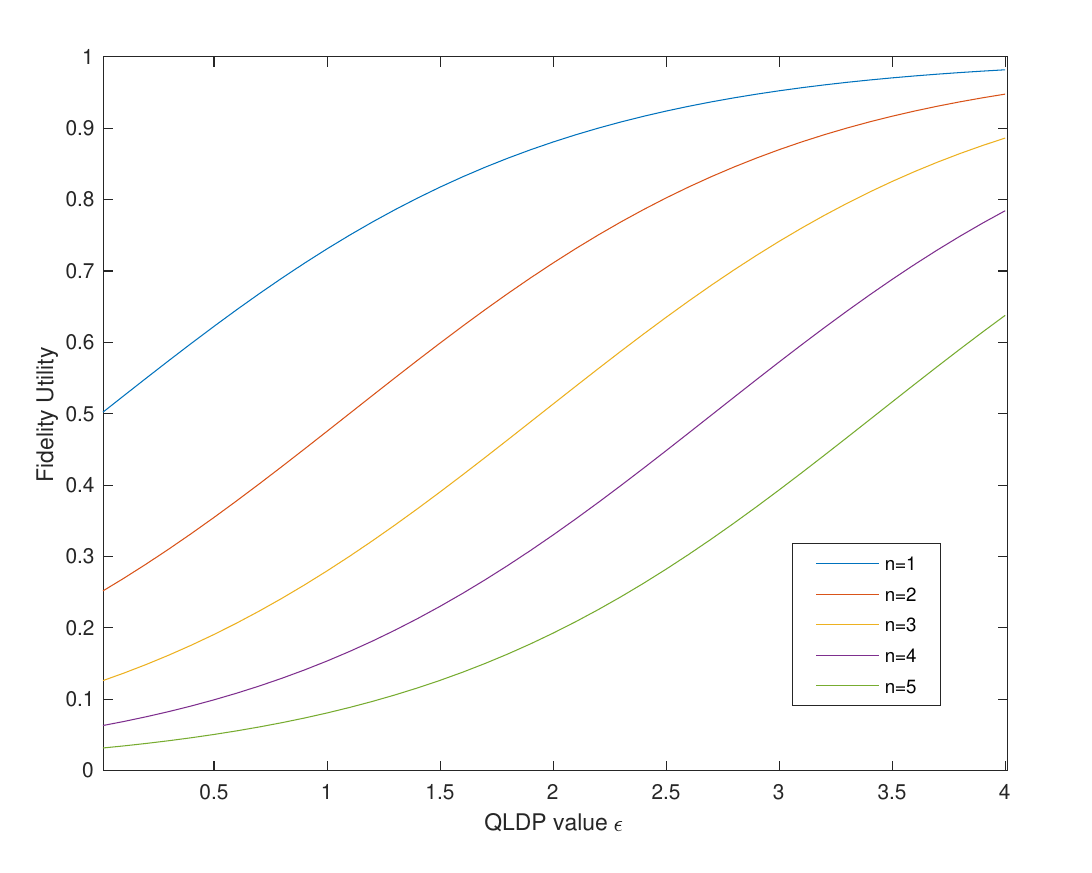}
    \caption{The trade-off between fidelity utility and QLDP of $n$-qubit  depolarizing quantum noise mechanism}
    \label{fig:tradeoff}
\end{figure}
\subsection{Trade-off}
Lemma~\ref{Lem:dep_utility_QLDP} showcases how adjusting the probability $p$ can lead to varying levels of the QLDP and fidelity utility for any $n$-qubit depolarizing noise. Theorem~\ref{thm:solution} establishes a trade-off between fidelity utility and QLDP values. By leveraging these observations, we investigate the trade-off concerning $n$-qubit depolarizing noise for values of $n$ from 1 to 5. The results are depicted in Fig.~\ref{fig:tradeoff}. This shows that the privacy-utility trade-off under depolarizing noise is strongly dependent on the number of qubits.

\begin{figure}[!htp]
    \centering
    \includegraphics[width=\linewidth]{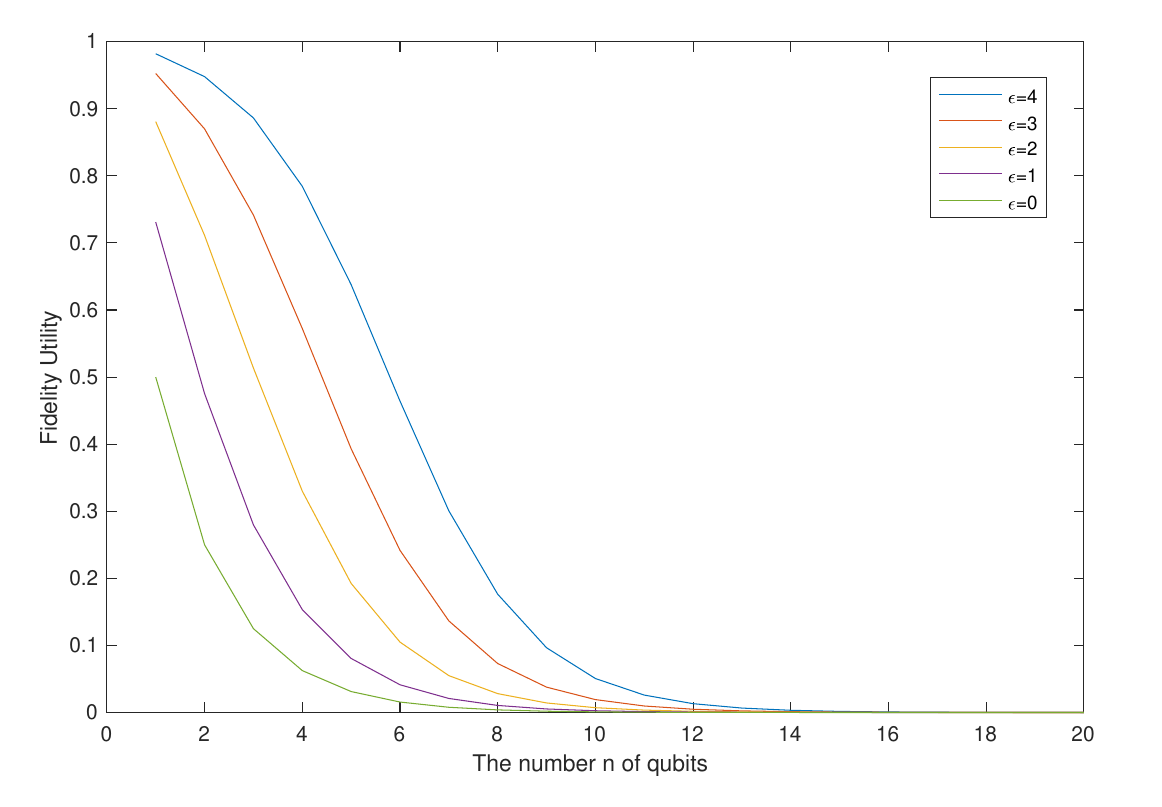}
    \caption{Fidelity utility decreasing with the number of qubits in depolarizing noise mechanisms}
    \label{fig:qubit_number}
\end{figure}

\subsection{Qubit Number}
As indicated by Theorem~\ref{thm:solution}, the fidelity utility of depolarizing noise diminishes exponentially as the number of qubits increases for a specified QLDP value $\epsilon$. This trend is visually represented in Fig.~\ref{fig:qubit_number} for $\epsilon=0,1,2,3,4$. It is apparent that applying depolarizing noise to a small number (2-4) of qubits is sufficient to maintain high utility for practical applications and achieve the desired QLDP value. However, in high-dimensional settings with larger numbers of qubits ($\geq  5$), enforcing strict $\epsilon$-QLDP can be overly restrictive, significantly degrading utility. To address this, it is necessary to consider relaxed notions of privacy such as $(\epsilon, \delta)$-differential privacy\cite{dwork2014algorithmic}, R{\'e}nyi differential privacy~\cite{mironov2017renyi}, or Pufferfish privacy~\cite{kifer2014pufferfish}, which can be adapted to the quantum local model to better balance privacy and utility in practical scenarios.

\section{Conclusion}
In this paper, we explore the optimal local differential privacy mechanism for quantum computing from a practical perspective. Initially, we demonstrate the feasibility of utilizing quantum noise mechanisms to uphold quantum local differential privacy (QLDP). Subsequently, we establish that the depolarizing noise mechanism, which acts as a quantum counterpart to the classical randomized response mechanism, stands out as the optimal choice for maximizing both fidelity and trace distance utility among all unital quantum mechanisms with a constant QLDP value. Furthermore, we analyze the trade-off between privacy and utility across a variety of quantum noise mechanisms, including both unital and non-unital types, using a combination of theoretical analysis and numerical simulations. These results underscore the effectiveness of quantum depolarizing noise within the QLDP framework. 

\section*{Acknowledgments}
Thanks to Yecheng Qin for pointing out errors in the earlier version of this paper and for providing the corrected results on the differential privacy and utility costs of the generalized amplitude damping noise mechanism. This work was partially supported by the Innovation Program for Quantum Science and Technology (Grant No. 2024ZD0300500), the Youth Innovation Promotion Association, Chinese Academy of Sciences (Grant No. 2023116),  the National Natural Science Foundation of China (Grant No. 62402485), the Young Elite Scientists Sponsorship Program, China Association for Science and Technology.

\bibliographystyle{ACM-Reference-Format}
\bibliography{main}

\newpage
\appendix
\section{Appendix}\label{Appendix}
\subsection{Proof of Lemma~\ref{lem:QLDP_LDP}}
{\it Proof.} This directly follows the definitions of QLDP in Definition~\ref{Def:QLDP} and classical LDP in Definition~\ref{Def:LDP}. 
\hfill $\Box$

\subsection{Proof of Theorem~\ref{Thm:LDP}}
{\it Proof.}
For any $\rho\not=\sigma\in\dh$, we can represent $\rho-\sigma=\Delta_{+}-\Delta_{-}$ as a decomposition into orthogonal positive and negative parts, where $\Delta_{\pm}> 0$ and $\Delta_{+}\Delta_{-}=0$. In the subsequent proof, we utilize the property $0<\tr(\Delta_{+})=\tr(\Delta_{-})\leq 1$ stemming from $\tr(\rho-\sigma)=0$ and $\rho-\sigma=\Delta_{+}-\Delta_{-}$, and thus $\frac{\Delta_{\pm}}{\tr(\Delta_{\pm})}$ are quantum states. The following derivation is based on these:
    \begin{equation*}
        \begin{aligned}                 
        &\tr[\e^\dagger(M_\s)(\rho-e^\epsilon\sigma)]\\
        =&\tr[\e^\dagger(M_\s)(\rho-\sigma+\sigma-e^\epsilon\sigma)]\\
        =&\tr[\e^\dagger(M_\s)(\Delta_{+}-\Delta_{-}+\sigma-e^\epsilon\sigma)] \\  
        =&\tr(\e^\dagger(M_\s)\Delta_{+})-\tr(\e^\dagger(M_\s)\Delta_{-})-(e^\epsilon-1)\tr(\e^\dagger(M_\s)\sigma) \\ 
        =&\tr(\Delta_{+})[\tr(\e^\dagger(M_\s)\frac{\Delta_{+}}{\tr(\Delta_{+})})-\tr(\e^\dagger(M_\s)\frac{\Delta_{-}}{\tr(\Delta_{-})})]\\
        &-(e^\epsilon-1)\tr(\e^\dagger(M_\s)\sigma) \\ 
        \leq &\tr(\Delta_{+})[\max_{\rho_1\in\dh}\tr(\e^\dagger(M_\s)\rho_1)-\min_{\rho_2\in\dh}\tr(\e^\dagger(M_\s)\rho_2)]\\
        &-(e^\epsilon-1)\min_{\rho_3\in\dh}\tr(\e^\dagger(M_\s)\rho_3)\\
        =&\tr(\Delta_{+})[\lambda_{\max}(\e^\dagger(M_\s))-\lambda_{\min}(\e^\dagger(M_\s))]-(e^\epsilon-1)\lambda_{\min}(\e^\dagger(M_\s))\\
        \leq &\lambda_{\max}(\e^\dagger(M_\s))-e^\epsilon\lambda_{\min}(\e^\dagger(M_\s))\\
        =&\bra{\psi^*}\e^\dagger(M_\s)\ket{\psi^*}-e^\epsilon\bra{\phi^*}\e^\dagger(M_\s)\ket{\phi^*}\\
        =&\tr[\e^\dagger(M_\s)(\psi^*-e^\epsilon\phi^*)]
        \end{aligned}
    \end{equation*}
    Here, $\ket{\psi^*}$ and $\ket{\phi^*}$ are the normalized eigenvectors (pure state) of $\e^\dagger(M_{\s})$ corresponding to the maximal and minimal eigenvalues, respectively. 
    Consequently, we have:
\[\max_{\rho,\sigma\in\dh}\tr[\e^\dagger(M_\s)(\rho-e^\epsilon\sigma)]=\lambda_{\max}(\e^\dagger(M_\s))-e^\epsilon\lambda_{\min}(\e^\dagger(M_\s))\]
This completes the proof based on the definition of QLDP provided in Definition~\ref{Def:QLDP}. 
\hfill $\Box$

\subsection{Proof of Lemma~\ref{Lem:QLDP_single}}
{\it Proof.}  For any measurement $\m_{M}$ with matrices $\{M,I-M\}$
\[\epsilon^*(\m_{M}\circ\e)=\max_{A\in\{I,M,I-M\}}\ln\frac{\lambda_{\max}[\e^\dagger(A)]}{\lambda_{\min}[\e^\dagger(A)]}\]
Noting that $\{I,M,I-M\}\subseteq \{M:0\leq M\leq I\}$, the set of all quantum measurement matrices. Combining that $\epsilon^*(\e)=\max_{0\leq M\leq I}\epsilon^*(\m_{M}\circ\e),$ we can complete the proof.
\hfill $\Box$

\subsection{Proof of Theorem~\ref{Thm:QLDP}}
{\it Proof.} By Lemma~\ref{Lem:QLDP_single}, $\e$ is $\epsilon$-QLDP if and only if for any measurement matrix $M$, 
\[0 \geq \lambda_{\max}(\e^\dagger(M))-e^\epsilon\lambda_{\min}(\e^\dagger(M)).\]
With the eigendecomposition of $M=\sum_{k}\lambda_k\psi_k$ and eigenvalues $\lambda_k\geq 0$, we have 
\[0 \geq \lambda_{\max}(\e^\dagger(\sum_{k}\lambda_k\psi_k))-e^\epsilon\lambda_{\min}(\e^\dagger(\sum_{k}\lambda_k\psi_k)).\]
By the linearity of $\e$, we have  
\[0 \geq \lambda_{\max}[\sum_{k}\lambda_k\e^\dagger(\psi_k)]-e^\epsilon\lambda_{\min}[\sum_{k}\lambda_k\e^\dagger(\psi_k)].\]
Noting the linear algebra fact that for any positive semi-definite matrices $A$ and $B$,
\begin{equation*}
    \begin{aligned}
        &\lambda_{\max}(A+B)\leq \lambda_{\max}(A)+\lambda_{\max}(B)\\
        &\lambda_{\min}(A+B)\geq \lambda_{\min}(A)+\lambda_{\min}(B).
    \end{aligned}
\end{equation*}
Applying these to the summation $\sum_{k}\lambda_k\e^\dagger(\psi_k)$ of positive semi-definite matrices $\e^\dagger(\psi_k)$, we obtain
\begin{equation*}
   \begin{aligned}
       &\lambda_{\max}[\sum_{k}\lambda_k\e^\dagger(\psi_k)]-e^\epsilon\lambda_{\min}[\sum_{k}\lambda_k\e^\dagger(\psi_k)]\\ 
  \leq& \sum_{k}\lambda_k[\lambda_{\max}(\e^\dagger(\psi_k))-e^\epsilon\lambda_{\min}(\e^\dagger(\psi_k))]\\
  \leq&(\sum_{j}\lambda_j)\max_{k}[\lambda_{\max}(\e^\dagger(\psi_k))-e^\epsilon\lambda_{\min}(\e^\dagger(\psi_k))].
   \end{aligned} 
\end{equation*}
Therefore, by noting that each $\psi_k$ is also a measurement matrix, we can claim that 
\[0 \geq \lambda_{\max}(\e^\dagger(M))-e^\epsilon\lambda_{\min}(\e^\dagger(M)) \quad\forall 0\leq M\leq I.\]
if and only if 
\[0 \geq \lambda_{\max}(\e^\dagger(\psi))-e^\epsilon\lambda_{\min}(\e^\dagger(\psi)) \quad\forall  \ket{\psi}\in\h.\]
This completes the proof and we have
\[\epsilon^*(\e)=\max_{\ket{\psi}\in\h}\ln\frac{\lambda_{\max}(\e^\dagger(\psi))}{\lambda_{\min}(\e^\dagger(\psi))}.\]
\hfill $\Box$


\subsection{Proof of Theorem~\ref{Lem:QLDP_checking}}
{\it Proof.} We first prove that $\epsilon^*(\e)<\infty$ if and only the first condition is satisfied. 

Initially, according to Theorem~\ref{Thm:QLDP}, it is evident that $\epsilon^*(\e)<\infty$ if and only if $\lambda_{\min}(\e^\dagger(\psi))>0$ for any $\ket{\psi} \in\h$. This condition is equivalent to stating that $\e^\dagger(\psi)>0$ holds true for any $\ket{\psi} \in\h$ based on the eigendecomposition of $\e^\dagger(\psi)$ with $\lambda_{\min}(\e^\dagger(\psi))$ being the minimum eigenvalue. Considering the linearity of $\e^\dagger$, the condition $\e^\dagger(\psi)>0$ for any $\ket{\psi} \in\h$ is equivalent to $\e^\dagger(\rho)>0$ for all $\rho\in\dh$. Such an $\e^\dagger$ is specifically referred to as strictly positive, which can be validated by examining whether the \emph{Choi-Jamiolkowski matrix} $\e^\dagger\otimes \i(\ketbra{\Omega}{\Omega})$ of $\e^\dagger$ is strictly positive, i.e., $\e^\dagger\otimes \i(\ketbra{\Omega}{\Omega})>0$, as detailed in \cite[Theorem 6.8]{wolf2012quantum}.

For a quantum mechanism $\e$ with Kraus matrices $\{E_k\}_{k=0}^{K-1}$, the computation of $\e^\dagger\otimes \i(\ketbra{\Omega}{\Omega})$ requires $\mathcal{O}(N^6K)$ operations, as shown below:

\[\e^\dagger\otimes \i(\ketbra{\Omega}{\Omega})=\sum_{k=0}^{K-1}(E_k^\dagger\otimes I)\ketbra{\Omega}{\Omega}(E_k\otimes I).\]

The complexity arises from the fact that there are $K$ summations and each of these involves two $N^2$-by-$N^2$ matrix multiplications with a computation cost of $\bigO{N^6K}$. Consequently, verifying if $\e^\dagger\otimes \i(\ketbra{\Omega}{\Omega}) > 0$ can be achieved by computing its eigenvalues, incurring a computational cost of $\mathcal{O}(N^6)$. Thus, the total complexity is $\mathcal{O}(N^6K).$

We proceed to demonstrate the equivalence of the two conditions. Initially, we observe that $\e^\dagger\otimes \i(\ketbra{\Omega}{\Omega}) > 0$ is the same as stating that the set formed by $\{(E_k^\dagger\otimes I)\ket{\Omega}\}$ spans the entire vector space $\h\otimes\h$. This equivalence arises from the fact that the mapping $E\rightarrow (E\otimes I)\ket{\Omega}$ establishes a linear one-to-one relationship between the matrix space in $\h$ and vector space $\h\otimes\h$. Addressing the issue of complexity to check the second condition, this leads to determining the rank $r$ of the linear space generated by $\{E_k\}_{k=0}^{K-1}$, achievable in $\bigO{N^6K}$. Consequently, $r=N^2$ holds if and only if the second condition is met.
\hfill $\Box$
\subsection{Proof of Theorem~\ref{Exa:noise_cost}}

{\it Proof.} According to Corollary~\ref{cor:QLDP_effectiveness}, we can verify the instances where quantum noises do not result in a finite QLDP value as claimed. We only need to calculate $\epsilon^*(\n_{\dep})$ for $0\leq p<1$ and $\epsilon^*(\n_{\gad})$ for $0<q<1$ and $0<r\leq 1$ in the following.

Recall that for any state $\ket{\psi}$, the following relation holds: 
\[\n_{\dep}(\psi) = p\psi + (1-p)\frac{I}{2}.\]
It can be observed that $\ket{\psi}$ and $\ket{\psi^{\bot}}$ represent the normalized eigenvectors associated with eigenvalues $(1+p)/2$ and $(1-p)/2$, respectively. Here, $\ket{\psi^{\bot}}$ represent a quantum state orthogonally to $\ket{\psi}$, i.e., $\braket{\psi}{\psi^{\bot}}=0$. Consequently, for $0\leq p<1$, we derive $\epsilon^*(\n_{\dep}) = \ln\frac{1+p}{1-p}$ based on Theorem~\ref{Thm:QLDP}.

To determine $\epsilon^*(\n_{\gad})$, according to Theorem~\ref{Thm:QLDP} again, our task involves solving the optimization problem $\max_{\psi}\ln{\frac{\lambda_{\max}(\n^\dagger_{\gad}(\psi))}{\lambda_{\min}(\n^\dagger_{\gad}(\psi))}}$. Initially, we represent a 1-qubit state $\ket{\psi}$ with two parameters $a$ and $b$ as \[\ket{\psi}=\begin{bmatrix}
    a\\
    b
\end{bmatrix} \text{ such that } aa^*+bb^*=1.\] 
To find the eigenvalues of $\n^\dagger_{\gad}(\psi)$, we start by computing the characteristic polynomial $\text{det}[\lambda I-\n^\dagger_{\gad}(\psi)]$ and obtain 

\begin{equation*}
\begin{aligned}
    &\text{det}[\lambda I-\n^\dagger_{\gad}(\psi)]\\
    =&\{\lambda-\frac{1}{2}[1+r(2q-1)(2x-1)]\}^2\\
    &-\frac{1}{4}[(1-r)^2(2x-1)^2+4(1-r)(1-x)x]
\end{aligned}   
\end{equation*}   
where $x=aa^*$.
By solving $\text{det}[\lambda I-\n^\dagger_{\gad}(\psi)]=0$, we find the eigenvalues of $\n^\dagger_{\gad}(\psi)$  being
\[\frac{1+r(2q-1)(2x-1)\pm \sqrt{(1-r)^2(2x-1)^2+4(1-r)(1-x)x}}{2}.\]
Then we have that $\ln{\epsilon^*(\n^\dagger_{\gad})}$ is 
\[{\max_{0\leq x\leq 1}\frac{1+r(2q-1)(2x-1)+ \sqrt{(1-r)^2(2x-1)^2+4(1-r)(1-x)x}}{1+r(2q-1)(2x-1)- \sqrt{(1-r)^2(2x-1)^2+4(1-r)(1-x)x}}}\]
The above value is achieved at $x=1-q$, resulting in the optimal value
\[\frac{1-r(1-2q)^2+\sqrt{(1-r)[1-r(1-2q)^2]}}{1-r(1-2q)^2-\sqrt{(1-r)[1-r(1-2q)^2]}}.\]
Hence, we conclude that 
   \[\epsilon^*(\n_{\gad})=\ln(\frac{\sqrt{1-rh^2}+\sqrt{1-r}}{\sqrt{1-rh^2}-\sqrt{1-r}})\]
        where $h=|1-2q|$.
    \hfill $\Box$

\subsection{Proof of Theorem~\ref{Exa:utility}}
{\it Proof.} First, we deal with $\n_\text{XF}, \n_\text{YF}$ and $\n_\text{ZF}$. They share the following same structure for their corresponding to super-operators:
\[\n_\text{UF}(\rho)=p\rho+(1-p)U\rho U^\dagger.\]
Then, the fidelity utility is
\[F(\n_\text{UF})=p+\min_{\ket{\psi}\in\h}|\bra{\psi}U\ket{\psi}|^2.\]

$\bra{\psi}U\ket{\psi}=0$ can be obtained by $\ket{\psi}=\ket{0}$ for $U=X$, $\ket{\psi}=\ket{0}$ for $U=Y$, and $\ket{\psi}=\frac{\ket{0}+\ket{1}}{\sqrt{2}}$ for $U=Z$. Thus $F(\n_\text{UF})=p$ for $U\in\{X,Y,Z\}$. 
The anti-trace distance utility is
\[T(\n_\text{UF})=(1-p)\max_{\ket{\psi}\in\h}\frac{1}{2}\tr(|U\psi U-\psi|)\]
$\frac{1}{2}\tr(|U\psi U-\psi|)=1$ can be obtained by $\ket{\psi}=\ket{0}$ for $U=X$,$\ket{\psi}=\ket{0}$ for $U=Y$, and $\ket{\psi}=\frac{\ket{0}+\ket{1}}{\sqrt{2}}$ for $U=Z$. Therefore, $\hat{T}(\n_{UF})=p$ for $U\in\{X,Y,Z\}$.

Now we consider the utilities of depolarizing noise. Recall one form of the noise is 
\[\n_{\dep}(\rho)=p\rho+(1-p)\frac{I}{2}.\]
Then
\[F(\n_{\dep})=\frac{1+p}{2}\]
\[T(\n_{\dep})=\frac{1-p}{2}\Rightarrow \hat{T}(\n_{\dep})=\frac{1+p}{2}.\]

Now we move to deal with damping noise mechanisms by parameterizing pure state $\ket{\psi}$.

Any 1-qubit $\ket{\psi}$ can be represented by two parameters $a,b$ as \[\ket{\psi}=\begin{bmatrix}
    a\\
    b
\end{bmatrix} \text{ with } aa^*+bb^*=1.\] 
Then we compute fidelity and anti-trace distance utilities by searching on such  $a$ and $b$.
\[\psi=\ketbra{\psi}{\psi}=\begin{bmatrix}
    aa^*& ab^*\\
    a^*b& bb^*
\end{bmatrix}.\]
For phase damping $\n_\text{PD}$,
\[\n_\text{PD}(\psi)=\begin{bmatrix}
    aa^*& \sqrt{1-r}ab^*\\
    \sqrt{1-r}a^*b& bb^*
\end{bmatrix}.\]
Then we have the eigenvalues of $\n_\text{PD}(\psi)-\psi$ are:
\[\lambda[\n_\text{PD}(\psi)-\psi]=\pm \sqrt{(1-\sqrt{1-r})^2aa^*bb^*}\]
 By $aa^*+bb^*=1$, we have the maximal value of $\sqrt{(1-\sqrt{1-r})^2aa^*bb^*}$ is   $\frac{1-\sqrt{1-r}}{2}$ for $aa^*=\frac{1}{2}$. Thus $\hat{T}(\n)=\frac{1+\sqrt{1-r}}{2}$.

On the other hand, $F(\n_\text{PD}(\psi),\psi)$ is
\[\bra{\psi}\n_\text{PD}(\psi)\ket{\psi}=(aa^*)^2+2\sqrt{1-r}aa^*bb^*+(bb^*)^2.\]
Let $x=aa^*$ with $0\leq x\leq 1$ and then $bb^*=1-x$ as $\braket{\psi}{\psi}=aa^*+bb^*=1$. Then the above equation is
\[F(\n_\text{PD}(\psi),\psi)=2(1-\sqrt{1-r})x^2-2(1-\sqrt{1-r})x+1.\]
Therefore, the minimal value of fidelity $F(\n_\text{PD}(\psi),\psi)$ is achieved at $x=\frac{1}{2}$ and thus the utility fidelity is $F(\n_\text{PD})=\frac{1+\sqrt{1-r}}{2}$.


For $\n_{\gad}$, 
\[\n_{\gad}(\psi)=\begin{bmatrix}
{(1-r)aa^*+rq}&\sqrt{1-r}ab^*\\
\sqrt{1-r}a^*b&raa^*+bb^*-rq
\end{bmatrix}.\]
Then we have the eigenvalues of $\n_{\gad}(\psi)-\psi$ are:
\[\lambda[\n_{\gad}(\psi)-\psi]=\pm\sqrt{(rq-raa^*)^2+(\sqrt{1-r}-1)^2aa^*bb^*}\]
Let $x=aa^*$ with $0\leq x\leq 1$ and then $bb^*=1-x$ as $\braket{\psi}{\psi}=aa^*+bb^*=1$. Then the magnitude of $\lambda[\n_{\gad}(\psi)-\psi]$ is
\[\sqrt{(r^2+2-2\sqrt{1-r}-r)x^2-(2qr^2+2-2\sqrt{1-r}-r)x+r^2q^2}\]
As $(r^2+2-2\sqrt{1-r}-r)\geq 0$, the maximal value of the above equation is achieved at $x=0$ or 1.
\begin{equation*}
   \begin{aligned}
       &\text{ For } x=1, |\lambda[\n_{\gad}(\psi)-\psi]|=(1-q)r.\\
       &\text{ For } x=0, |\lambda[\n_{\gad}(\psi)-\psi]|=qr.
   \end{aligned} 
\end{equation*}
Thus 
\[T(\n_{\gad})=\left\{
    \begin{aligned}
        &qr &\text{ if } q\geq \frac{1}{2}\\
        &(1-q)r &\text{ if } q< \frac{1}{2}
    \end{aligned}\right. .\]
Then
\[\hat{T}(\n_{\gad})=\left\{
    \begin{aligned}
        &1-qr &\text{ if } q\geq \frac{1}{2}\\
        &1-(1-q)r &\text{ if } q< \frac{1}{2}
    \end{aligned}\right. .\]
    
On the other hand, $F(\n_\text{GAD}(\psi),\psi)=\bra{\psi}\n_\text{GAD}(\psi)\ket{\psi}$ is
\[(1-r)(aa^*)^2+rqaa^*+(2\sqrt{1-r}+r)aa^*bb^*+(bb^*)^2-rqbb^*.\]
Let $x=aa^*$ with $0\leq x\leq 1$ and then $bb^*=1-x$ as $\braket{\psi}{\psi}=aa^*+bb^*=1$. Then the above equation is
\[2(1-r-\sqrt{1-r})x^2+(2rq+r+2\sqrt{1-r}-2)x+1-rq.\]
Therefore, the minimal value of fidelity $F(\n_\text{GAD}(\psi),\psi)$ is achieved at $x=0$ or 1 and thus the utility fidelity is 
\[F(\n_{\gad})=\left\{
    \begin{aligned}
        &1-qr &\text{ if } q\geq \frac{1}{2}\\
        &1-(1-q)r &\text{ if } q< \frac{1}{2}
    \end{aligned}\right. .\]
     \hfill $\Box$
\subsection{Proof of Theorem~\ref{Thm:tradeoff}}
    {\it Proof.} Let $D$ be the dimension of the state space $\h$ which $\n$ is working on. Then, by Theorem~\ref{Thm:QLDP}, we have 
\begin{equation*}
    \begin{aligned}   
    &e^{\epsilon^*(\n)}\\
    =&\max_{\ket{\psi}}\frac{\lambda_{\max}(\n^\dagger(\psi))}{\lambda_{\min}(\n^\dagger(\psi))}\\
    \geq&\max_{\ket{\psi}}\frac{(D-1)\lambda_{\max}(\n^\dagger(\psi))}{1-\lambda_{\max}(\n^\dagger(\psi))}\\
    =&\max_{\ket{\psi}}(D-1)(\frac{1}{1-\lambda_{\max}(\n^\dagger(\psi))}-1)\\
    \geq& \max_{\ket{\psi}}(D-1)(\frac{1}{1-\bra{\psi}\n^\dagger(\psi)\ket{\psi}}-1)\\
    =&(D-1)(\frac{1}{1-\max_{\psi}\bra{\psi}\n^\dagger(\psi)\ket{\psi}}-1)\\
    \geq&(D-1)(\frac{1}{1-F(\n)}-1).
    \end{aligned}
\end{equation*}
Then we have the inequality for fidelity  utility for $D=2^n$:
\[F(\n)\leq \frac{e^{\epsilon^*(\n)}}{e^{\epsilon^*(\n)}+2^n-1}.\]
In the above, the first inequality results from that for any $D$-by-$D$ positive semi-definite matrix $M$ with $\tr(M)=1$ (eigenvalues of $M$ are all non-negative and the summation of them is 1), we have 
\[\lambda_{\min}(M)\leq \frac{1-\lambda_{\max}(M)}{D-1}.\] 
The second inequality comes from \[\lambda_{\max}(\n^\dagger(\psi))=\max_{\phi}\bra{\phi}\n^\dagger(\psi)\ket{\phi}.\]
The third inequality results from
\[F(\n)=F(\n^\dagger)=\min_{\psi}\bra{\psi}\n^\dagger(\psi)\ket{\psi}\leq \max_{\psi}\bra{\psi}\n^\dagger(\psi)\ket{\psi}\]
And we observe that  the above two equalities are obtained if and only if 
   \[\n^\dagger(\psi)=\eta_{0}{\psi}+\sum_{1\leq k\leq D-1}\eta_{k}{\psi_{k}} \qquad\forall \ket{\psi}\in\h\]
   where the set $\{\ket{\psi},\ket{\psi_1},\ket{\psi_2},\ldots, \ket{\psi_{D-1}}\}$ of pure state are mutually orthogonal and $\eta_0\geq \eta_1=\eta_{2}=\cdots =\eta_{D-1}.$ That is
   \[\n^\dagger_{\eta}(\psi)=\eta{\psi}+\frac{1-\eta}{D-1}(I-{\psi}).\]
   where $\eta\geq \frac{1}{D}$ is a constant and  independent on $\psi$. 

  Furthermore,  by noting that $\t(\n)\geq 1-\f(\n)$, we can get the second inequality for anti-trace distance fidelity.
\hfill $\Box$

\subsection{Proof of Lemma~\ref{Lem:dep_utility_QLDP}}
{\it Proof.} This can be accomplished directly using Theorem~\ref{Thm:QLDP} and the definitions of fidelity and anti-trace distance utilities, similar to how we handle 1-qubit depolarizing noise previously.  \hfill $\Box$

\subsection{Proof of Theorem~\ref{thm:solution}}
{\it Proof.} It directly follows from Theorem~\ref{Thm:tradeoff} and Lemma~\ref{Lem:dep_utility_QLDP}. \hfill $\Box$

\end{document}